\documentclass[journal,10pt]{IEEEtran}
\IEEEoverridecommandlockouts
\usepackage{amssymb}
\usepackage{amsmath}
\usepackage{graphicx}
\usepackage{amsmath,amsthm}
\usepackage{float}
\usepackage{enumerate}
\usepackage{epstopdf}
\usepackage{dblfloatfix}
\usepackage{caption}
\usepackage{amsfonts}
\usepackage{fancyhdr}
\usepackage{bbm}
\usepackage{cases}
\usepackage{cite}
\usepackage{breqn}
\usepackage{multirow}
\usepackage{mathrsfs}
\usepackage{algpascal}
\usepackage{algc}
\usepackage{algorithmicx}
\usepackage[ruled]{algorithm}
\usepackage{algpseudocode}
\usepackage{graphics}
\usepackage{epsfig}
\usepackage{mathrsfs} % the form of set
\usepackage{dsfont}
\usepackage{tabularx}
\usepackage{color}
\usepackage{booktabs}
\usepackage{subfigure}

\usepackage{tikz}
\usetikzlibrary{shapes,arrows,positioning,calc}
\usepackage[colorlinks=true, linkcolor=blue, citecolor=blue, urlcolor=blue]{hyperref}

\newcommand{\tocaption}{%
\setlength{\abovecaptionskip}{0pt}%
\setlength{\belowcaptionskip}{10pt}%
\caption}

\floatname{algorithm}{Alg.}

\makeatletter
\newcommand{\multiline}[1]{%
  \begin{tabularx}{\dimexpr\linewidth-\ALG@thistlm}[t]{@{}X@{}}
    #1
  \end{tabularx}
}

\algdef{SE}[DOWHILE]{Do}{doWhile}{\algorithmicdo}[1]{\algorithmicwhile\ #1}%

\ifCLASSINFOpdf
 \else
\fi

\begin{document}
\theoremstyle{definition} % Set the style of Theorem
\newtheorem{theorem}{Theorem}[section]
\newtheorem{definition}[theorem]{Definition}
\newtheorem{lemma}[theorem]{Lemma}
\newtheorem{example}[theorem]{Example}
\newtheorem{Proposition}[theorem]{Proposition}
\newtheorem{Corollary}[theorem]{Corollary}
%\newtheorem{remark}[theorem]{Remark}

% \title{Fast Beam Prediction in mmWave Communications: A Memristor-Based Meta-Learning Framework}
% \title{Memristor-Based Meta-Learning for Fast Terahertz
% Beam Prediction in Non-Stationary Environments}
% \title{Fast mmWave
% Beam Prediction in Non-Stationary Environments: A Memristor-Based Meta-Learning Framework}
% \title{Fast mmWave Beam Prediction in Non-Stationary Environment using Memristor-Based Meta Learning}
% \title{Beam Prediction in Non-Stationary Environment using Memristor-Based Meta Learning}
% \title{Fast Beam Prediction in Non-Stationary Environment using Memristor-Based Meta Learning}
\title{Memristor-Based Lightweight Meta Learning for Beam Prediction in Non-Stationary Environments}

\author{
Yuwen Cao\textsuperscript{},~\IEEEmembership{Member, IEEE}, Tomoaki Ohtsuki\textsuperscript{},~\IEEEmembership{Senior Member, IEEE}, Setareh Maghsudi\textsuperscript{},~\IEEEmembership{Member, IEEE}, and Tony Q. S. Quek\textsuperscript{},~\IEEEmembership{Fellow, IEEE}
%
% \thanks{
% The work of Y. Cao was supported in part by the National Natural Science Foundation of China under Grant 62301143, and in part by Shanghai Sci-Tech Co-Research Program under Grant 25HB2702400. 
% The work of S. Maghsudi was supported by the Germany Federal Ministry of Research, Technology and Space (BMFTR) under Grant 16KIS2411.
% %
% The work of T. Q. S. Quek was supported in part by the National Research Foundation, Singapore and Infocomm Media Development Authority under its Communications and Connectivity Bridging Funding Initiative. Any opinions, findings and conclusions or recommendations expressed in this material are those of the author(s) and do not reflect the views of National Research Foundation, Singapore.
% This paper was presented in part at the IEEE Conference on Communications (ICC) 2025 \cite{11161557}. The
% associate editor coordinating the review of this article and approving it
% for publication was Prof. S. Mao. (\textit{Corresponding author:
% Tony Q. S. Quek.})
% }
\thanks{
This paper was presented in part at the IEEE Conference on Communications (ICC) 2025 \cite{11161557}.
Y. Cao is with the School of Information and Intelligent Science, Donghua University, Shanghai 201620, China  (e-mail: ywcao@dhu.edu.cn).
T. Ohtsuki is with the Department of Information and Computer Science, Keio University, Yokohama 223-8522, Japan (e-mail: ohtsuki@ics.keio.ac.jp).
S. Maghsudi is with the 
Department of Electrical Engineering and Information Technology, Ruhr-University Bochum, 44801 Bochum, Germany (e-mail: setareh.maghsudi@ruhr-uni-bochum.de).
T. Q. S. Quek is with the Singapore University of Technology and Design, Singapore 487372, and also with the Department of Electronic Engineering, Kyung Hee University, Yongin 17104, South Korea (e-mail: tonyquek@sutd.edu.sg).}
}

% \markboth{IEEE XX XX XX,~Vol.~XX, No.~XX, XXX~2024}
% {}

\maketitle

\begin{abstract}
Traditional machine learning techniques have achieved great success in improving data-rate performance and reducing latency in millimeter wave (mmWave) communications.
However, these methods still face two key challenges: (i) their
reliance on large-scale paired data for model training and tuning,
which limits performance gains and makes beam predictions
outdated, especially in multi-user mmWave systems with large
antenna arrays, and
%The performance gain brought by these methods is often limited by their reliance on large-scale paired
%data and the high training/tuning overhead, thus adversely affecting their applications in multi-user mmWave system installing large-scale antenna array. 
%Moreover, meta learning-based beamforming approaches can enable a swift adaption to unknown
%or nonstationary environments by training a small amount of data and thus obtaining a good pre-trained model. However, 
%
(ii) meta-learning (ML)-based beamforming
solutions are prone to overfitting when trained on a limited
number of tasks.
To address these challenges, we first propose a memristor-based meta-learning (M-ML) framework to expedite spatial and temporal domain beam prediction. Notably, the M-ML framework generates optimal
initialization parameters during the training phase, providing
a strong starting point for adapting to unknown environments
during the testing phase.  By leveraging memory to store key data, M-ML ensures the predicted beamforming vectors are well-suited to episodically dynamic channel distributions, even when testing and training environments do not align.
Afterwards, we propose a Gaussian noise-based regularized meta-learning framework to model the uncertainty in the training data and improve its stability and accuracy in complex environments.
Simulation results manifest that our approaches deliver high prediction accuracy in new environments, without relying on large datasets. Moreover, M-ML enhances the model's generalization ability and adaptability. 
\end{abstract}


\section{Introduction}
With the continuous advancement of technology, millimeter wave (mmWave) communication has emerged as a key technology for 5G communications \cite{Qualcomm2022,11145277,10361836}. With the continuous development of the 6G technology, there are higher requirements in terms of the data rate, reliability and delay \cite{8869705,7876965,10036372}. In this context,  accurately predicting mmWave beams between base stations (BSs) and mobile users has attracted significant research interest.
To reduce the training overhead incurred for beam prediction, the beamforming vectors can be transformed into multiple low-dimensional vectors, and the optimization of the beamforming vectors can usually be done by solving a weighted sum rate (WSR) maximization problem under total power constraints. However, WSR maximization is non-convex and NP-hard in practice \cite{4712693}.

To tackle this challenge, an exhaustive search algorithm, such as sector scanning, has been developed to identify the best beam \cite{Steinmetzer2017CompressiveMS,5756489}. Specifically, in \cite{Steinmetzer2017CompressiveMS}, two stations send detection frames to each other to measure the received signal strength of each sector, and then the sector with the highest signal strength is selected to establish a connection. 
Since the complexity of this algorithm increases linearly along with the number of predefined sectors, such algorithm is valid only for a limited number of sectors. 
To overcome the limitations in terms of scanning speed involved in traditional methods, \cite{9764610} uses machine learning for beam prediction, i.e., the beam prediction is accelerated by pre-training the mapping function, which avoids the exhaustive search incurred by traditional methods \cite{Steinmetzer2017CompressiveMS,5756489}. 
The authors in \cite{9314253} propose a low-overhead beam selection scheme via precisely estimating the received power
by exploiting the deep neural
network (DNN) and super-resolution techniques.
Reference \cite{10096315} proposes a model-agnostic \textit{meta-learning} (MAML), as a model-independent meta-learning (ML) algorithm, to train an initialized model parameter so that it can achieve good generalization on a new task by using iterative gradient updates.
Notably, the idea of a model-independent ML is presented in reference\cite{Finn2017ModelAgnosticMF} for the first time, which can be used for any model and learning task trained using gradient descent (GD).
% The goal of model-independent ML is to initialize parameters and obtain a pre-trained model by training on multiple tasks using only a small number of samples during testing  
The goal of model-independent ML is to obtain a pre-trained model, which can then be fine-tuned with only a small number of samples during testing to learn and adapt quickly to new tasks. Hence, model-independent ML can help to overcome the bottlenecks involved in traditional machine learning techniques, thereby can be utilized to make accurate predictions with low overhead.
Furthermore, Yuan \textit{et al.} 
in \cite{9257198} develops an ML-based approach to solve the task mismatch problem by training in a multi-task environment to obtain a generalized initialization model, as well as to achieve accurate beam prediction.
Compared with traditional machine learning, the initialized model obtained through ML training can perform beam prediction in unseen scenarios and the accuracy obtained is higher than traditional machine learning algorithms.
However, when encountering episodically dynamic non-stationary environments for training, we observe that the aforementioned ML-based solutions (i) are prone to \textit{meta-overfitting} when handling a limited number of training tasks; (ii) may face the risk of \textit{catastrophic forgetting}. Therefore, it is critical to come up with new  mmWave beam prediction solution to tackle the abovementioned challenging problems.

%  As a model-independent meta-learning algorithm, model-agnostic meta-learning (MAML) is proposed to train an initialised model parameter so that it can achieve good generalisation on a new task by training one or more gradient updates \cite{10096315}. Reference 
% \cite{9257198} proposes a meta-learning approach is proposed to solve the task mismatch problem by training in a multi-task environment to obtain a generalised initialisation model and, at the same time, achieve fast beam prediction when a vehicle enters a new environment. Compared with traditional machine learning, the initialised model obtained through meta-learning training can perform fast beam prediction in unseen scenarios and the accuracy obtained is much better than traditional machine learning.
 
The core goal of ML \cite{Thrun1998,Vinyals2016MatchingNF,Tseng2020CrossDomainFC,10847789,9954418,10418866,9413598,10477590}, also known as \textit{learning to learn}, is to enable models to quickly master new tasks using a small number of samples by exploiting the experience of previous tasks. 
% Predictions are made by referring to features encoded in the input data and training examples in the general metric space, and this approach is groundbreaking in the field of machine learning because it gets rid of the dependence on large amounts of 
%  training data, \textcolor{red}{thus reducing the cost incurred by data collection and labeling.} 
In addition, ML significantly reduces the time and computational resources required for the system to adapt to new situations, which is especially important in scenarios where data is scarce or rapidly updated.
It is noteworthy that the ML technique as well as its variant, i.e., few-shot ML technique have been presented in references \cite{Thrun1998,Vinyals2016MatchingNF,Tseng2020CrossDomainFC}. In addition, scenario-adaptive ML techniques for beam prediction have been investigated in \cite{10847789} for mmWave communication systems and in \cite{9954418} for dual-band systems. Besides, in \cite{10418866,9413598,10477590}, the authors developed advanced meta reinforcement learning algorithms for reconfigurable intelligent surface (RIS)-assisted wireless communications.
% \textcolor{red}{However, ML models still face the risk of meta-overfitting when dealing with a limited number of training tasks. 
% % Meta-overfitting occurs when a model overfits the training data to the extent that it loses its ability to predict new data.} 
% Usually, Meta-overfitting occurs in ML when the number of training tasks is insufficient to support the model to generalize the features. For instance, in a classification task with only few categories, the model may learn features specific to those categories while ignoring broader concepts, leading to performance degradation when encountering new categories\cite{Kim2018BayesianMM}.
% To deal with the overfitting challenge in supervised learning, researchers have invested a great deal of work focusing on the development of models that can learn a single task, such as a system that recognizes a fixed class in both the training and testing sessions. In this process, the dropout technique is widely used in the training phase of deep neural networks to reduce the model's dependence on training data by randomly setting some network activations to zero \cite{Tseng2020RegularizingMV}.
%

Numerous advanced techniques
attempting to enhance the traditional dropout methods have been proposed in
\cite{Tompson2014EfficientOL, Ghiasi2018DropBlockAR, pmlr-v28-wan13, Larsson2016FractalNetUN}
through introducing structural noise or adjusting the timing of the dropout. However, these improvements are beneficial only for single-task training models and may be infeasible for ML environments, which require models to migrate knowledge and experience among tasks.
Furthermore, to provide more flexible regularization, the Gaussian dropout method is proposed in the reference\cite{tseng2020regularizing}, which relaxes the binary randomness restriction of the traditional dropout methods by introducing Gaussian-distributed noise into the activation function. As a result, this will add more variability to the training process of the model. Therefore, 
% the Gaussian dropout approach not only maintains the original antioverfitting effect of dropout, but also provides an additional enhancement to the generalization ability of the network through the continuous noise distribution.
the Gaussian dropout approach not only retains the original anti-overfitting effect of dropout but also enhances the generalization ability of the network further through the continuous noise distribution.

On the other hand, in dynamic environments, when a vehicle enters a new environment for training, the trained neural networks may overwrite or forget old knowledge, i.e., a phenomenon known as catastrophic forgetting. Continuous learning is an efficient solution to the catastrophic forgetting problem, and its core idea is that during training, new tasks can be learnt quickly on the basis of existing knowledge without forgetting the previously learnt knowledge. A continuous learning method based on ML representations is proposed in \cite{Javed2019MetaLearningRF}, in which the objective is to minimize the catastrophic interference based on ML representations. Such method accelerates future learning, and is robust to forgetting under the online update of continuous learning. The problem of resource optimization for wireless communication systems in dynamic environments is presented in \cite{9926157,9682542,10622978,10478627,10173560}. Subsequently, a solution for fast adaptation to the environment is developed in conjunction with deep learning networks, where catastrophic forgetting is avoided through the use of a memory block to remember the data from the previous moments and train it together with the data from the next moments. 
% This approach allows for online learning and memory mechanisms that enable models to quickly adapt to new environments and tasks while maintaining superior performance on previous tasks.
Such approach allows for online learning and data depository mechanism that enables model to quickly adapt to new environments and tasks while maintaining superior performance on previous tasks.

\subsection{Motivations and Our Contributions}
%
% The problem of resource optimization for wireless communication systems in dynamic environments is presented\cite{9926157}\cite{9682542}, and a solution for fast adaptation to the environment is considered in conjunction with deep learning networks, where catastrophic forgetting is avoided through the use of a memory block to remember the data from the previous moments and train it with the data from the next moments. The approach allows for online learning and memory mechanisms that enable models to quickly adapt to new environments and tasks while maintaining performance on previous tasks.
Based on the aforementioned statements and analysis, we observe that with a small amount of data, traditional ML models are prone to meta-overfitting. In addition, traditional ML pre-trained models usually need to be tested in a new environment after completing training to verify their generalization ability. However, since the new environment may be quite different from the training environment, it is often difficult for the model to achieve optimal performance at the initial stage, and thus it takes a longer time to adapt to a new environment, particularly for multi-user mmWave systems with large
antenna arrays. Besides, the current existing ML techniques face the risk of catastrophic forgetting, thus rendering severe performance degradation when employing these existing ML techniques for beam predictions in dynamic environments using a small
amount of collected data. 

%-------------------------->Table: Symbol Description
\begin{table*}
\captionsetup{format=plain, singlelinecheck=off, labelsep=newline, font={small,sc}}
\captionsetup{justification=centering}
\tocaption{\newline \footnotesize{List of Most Commonly-Used Variables}}
  \centering
  %\footnotesize
  \scriptsize
  \begin{tabular}{|c|c|}
    \hline
    \textbf{Symbols} & \textbf{Definitions/Explanations}
       \\
       \hline
       \hline
      $N$ &  {The number of antennas at the BS.} \\
       \hline
      $K$ &  {The number of single-antenna users.} \\
      \hline
       $P$ & {The maximum downlink transmission power at the BS.}\\
       \hline
      $\mathbf{v}_{k}$ & {The transmit beamforming vector for user $k\in\{1,\ldots,K\}$.}\\
      \hline
      $\alpha_k$ & {The system weight of user $k$, which is a value controlled by the communication system.}\\
       \hline
      $\mathcal{D}_{s}$; $\mathcal{D}_{q}$ & {The training dataset (i.e., the support dataset) and the validation dataset (i.e., the query dataset), respectively.} \\
       \hline
       $Loss_{\mathcal{D}_{s}(\kappa)}\left(\theta_{\kappa}\right)$; $Loss_{\mathcal{D}_{q}}\left(\psi\right)$ & {The loss function for task $\kappa$ in the inner loop, and the loss function for tasks in the outer loop, respectively.} \\
      \hline
      $M$; $\mathcal{M}_{t}$
      & {The total memory size, and the memory set comprising of the latest test data and the selected representative data at time instance $t$.} \\
       \hline
      $\Omega_{Dt}$; $\Omega_{Mt}$ & {The index set of the latest test data, and the index set of selected representative samples, respectively.} \\
      \hline
      $\hat{\Omega}_{M_t}$; $\hat{\Phi}_{M_t}$ & 
      {Candidate representative samples $\mathcal{M}_t$ selected via the loss sensitivity and
      $k$-NN based memory updating mechanisms, respectively.}\\
       \hline
       $\lambda_{k}$ & {The virtual power allocation vector for user $k$, satisfying the constraint $\sum_{k=1}^{K} \lambda_{k} = P$.} \\
       \hline 
       $w_{k}$; $u_{k}$; $\mu$ & {The low-dimensional components decomposed from the beamforming matrix $\mathbf{V} = [\mathbf{v}_1, \ldots, \mathbf{v}_{K}]$.}\\
       \hline
       $\theta_u$; $\theta_w$; $\theta_{\mu}$ & {The inner model parameters w.r.t. the predicted low-dimensional components $\mathbf{u}$, $\mathbf{w}$, $\mu$, respectively.} \\
       \hline
       $\varphi_u$; $\varphi_w$; $\varphi_{\mu}$ & {The outer model parameters w.r.t. the  predicted low-dimensional components $\mathbf{u}$, $\mathbf{w}$, $\mu$, respectively.} \\
       \hline
       $P_g$ & {The Gaussian Gradient dropout probability.}\\
       \hline
       $m$ & {The shape parameter of the Nakagami distribution.} \\
       \hline
  \end{tabular}
  \label{Tb:Symbols}
\end{table*}
%-------------------->

Against this background, we study the mmWave beam prediction in multiple-input single-output
(MISO) over downlink broadcast channel (MISO-BC) by exploiting the MAML framework. Specifically, we first decompose the high-dimensional beamforming matrix into low-dimensional components based on the weighted minimum mean squared error (WMMSE) algorithm, followed by formulating a low-dimensional component prediction problem. To address this challenging problem and based on the memristor concept, we propose a new meta-learning framework, which is called as memristor-based meta-learning (M-ML), to expedite
spatial and temporal domain beam prediction.\footnote{Notice that memristor is a nonlinear component with the property of enhancing the model’s ability of dynamically learning and memorizing the features of new environments.} Afterwards, we use these predicted components to precisely recover the beamformer. Notably, M-ML exploits memory to store key data for facilitating a swift adaption to episodically dynamic channel environment, thus yielding superior performance. 
Note that, we conceive the M-ML framework by integrating the storage property of memristors into meta learning for the first time, to the best of our knowledge, attempting to address the \textit{catastrophic forgetting} encountered in beam prediction in non-stationary environments.
Besides, we develop a Gaussian noise-based regularized meta-learning (GN-RML) framework to model the uncertainty in the training data and improve its stability and accuracy in complex environments.
Also note that, the main contributions of this work can
be summed up as follows:
\begin{itemize}
\item [$\bullet$] In this paper, we formulate a joint beam prediction and memory set selection problem in MISO-BC for predicting mmWave beam with high accuracy while improving the model’s generalization ability and adaptability in new environments.
\item[$\bullet$] We propose a M-ML framework for predicting mmWave beam. The proposed framework can enable high sum rate by prioritizing and sampling these data points with the high loss in memory set, followed by combining these samples with the new training data to participate in model updating and optimization in dynamic environments. Furthermore, we design a $k$-nearest neighbors ($k$-NN) based memory updating strategy to classify the channel distributions from the data set. Subsequently, we select these matched channel groups together with the latest data capturing the vital characteristics and patterns of the new environment for outer model testing.   
\item[$\bullet$] In addition, based on the traditional beam prediction, a regularization method is introduced to improve the performance of the model and prevent meta-overfitting phenomenon by comparing the weight regularization with the regularization method that involves Gaussian noise.
\item[$\bullet$] Simulations are carried out in practical communication scenario which is episodically dynamic. Comparing M-ML with the state-of-the-art WMMSE method \cite{5756489}, unsupervised training, and MAML \cite{10096315}, we deduce that adding memory to store data dynamically can improve the performance of the model very well in the testing process.
\end{itemize}

The rest of this paper is organized as follows. Section II introduces the system
model and problem formulation. Section III presents the proposed M-ML aided beam prediction framework as well as the memory set selection mechanism to address the above challenging problem with low overhead. Subsequently, the GN-RML-assisted beam prediction framework is introduced in 
Section IV. Finally, simulation result and the concluding remarks are provided in Sections V and VI, respectively.

\textit{Notations}: The boldface lower-case letters and capital letters are used to represent the column vectors and matrices, respectively. By $(\cdot)^{T}$, $(\cdot)^{-1}$, $(\cdot)^{H}$, $|\mathcal{M}|$, and $\mathrm{Tr}(\cdot)$, we mean the transpose, the inverse, the Hermitian conjugate transpose, the cardinality of set $\mathcal{M}$, the trace operation of a square matrix, respectively.
$|\cdot|^2$ is the square of the modulus used to compute the vector,
while $\|\cdot\|_{p}$ means the $l_p$-norm operator.
$\odot$ is an element-level product (Hadamard product) operation for accurate element-by-element multiplication. In addition, $\mathcal{CN}(\mu, \boldsymbol{\Sigma})$ denotes the multivariate complex Gaussian
distribution with mean $\mu$ and covariance matrix $\boldsymbol{\Sigma}$. $\mathbb{E}(\cdot)$ accounts for the expectation of the argument.
$\nabla_{x}$ means the partial derivative of the loss function with respect to (w.r.t.) the involved parameter $x$. Besides, $\mathbf{I}_{M}$ means an identity matrix of size $M\times M$.
Note that Table \ref{Tb:Symbols} summarizes the frequently-used variables in the order of appearance in the paper.

\section{System Model and Problem Formulation}
\subsection{MISO-BC Transmission System}
\begin{figure}[tp]
\setlength{\abovecaptionskip}{-0.1cm}
\setlength{\belowcaptionskip}{-0.5cm}
    \begin{center}
    \includegraphics[width=4.0cm]{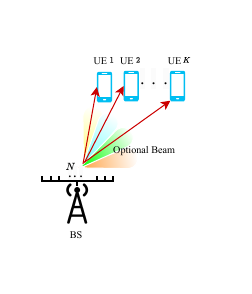}
    \end{center}
   \caption{Schematic diagram for the system model.}\label{fig:01}
\end{figure}

Consider a MISO-BC transmission system whereby the BS with $N$ antennas serves $K$ single-antenna users (UEs), as shown in Fig. \ref{fig:01}. Then, the signal observed at the $k$th UE can be expressed as
\begin{equation}
{{y}}_k=\mathbf{{h}}_k^H\mathbf{{v}}_k{s}_k+{\sum\nolimits_{i =1,i\neq k}^{K}\mathbf{{h}}_k^H\mathbf{{v}}_i}{s}_i+{n}_k,
\end{equation}
where 
$\mathbf{h}_k\sim\mathcal{CN}(0,\mathbf{I}_N)$ denotes the channel distribution between the BS and the $k$th UE.
$\mathbf{v}_k\in\mathbb{C}^{N_{}}$ is the transmit beamforming vector for the UE $k$.
${n}_k \sim\mathcal{CN}(0,\sigma^{2})$ represents the circularly symmetric Gaussian noise with zero mean and variance $\sigma^{2}$.
Let $s_k\sim$ $\mathcal{CN}(0,1)$ be the independent data symbol transmitted to the $k$th UE.  

The instantaneous signal-to-interference-plus-noise-ratio (SINR) received at the UE $k$ is given by
\begin{equation}
\text{SINR}_k=
\frac{|\mathbf{h}_k^H\mathbf{v}_k|^{2}}{\sigma^{2} + \sum_{i=1,i\neq k}^K{|\mathbf{h}_k^H\mathbf{v}_i|}^{2}}.
\end{equation}
Let $\textit{P}$ be the maximum transmission power and $\alpha_k$ denote the system weight of UE $k$. For a given estimated channel distribution, the instantaneous WSR maximization problem involved in the MISO-BC transmission system under the total transmit power constraints can then be expressed as:
\begin{equation}
\begin{aligned}
\label{eq:003}
&\mathop{\text{maximize}}_{\mathbf{V}}\sum\nolimits_{k=1}^{K}{\alpha}_k\text{log}_2{(1+\text{SINR}_k)}\\
&s.t. \; { \mathrm{Tr}(\mathbf{VV}^H)\leq}{P},
\end{aligned}
\end{equation}
where $\mathbf{V} = [\mathbf{v}_{1}, \ldots,\mathbf{v}_{K}]\in\mathbb{C}^{N\times K}$ is the transmit beamforming matrix for $K$ UEs at each time instance. Note that the optimization problem (3) is difficult to solve for the following reasons: i) the problem (3) is non-convex and ii) using the WMMSE algorithm \cite{5756489} involves the time-consuming iteration process, thus rendering the estimated beamforming weights outdated as the small-scale fading varies in the order of milliseconds. Besides, it is pointed out by reference \cite{6832894} that, the optimum solution to the instantaneous WSR maximization problem (\ref{eq:003}) follows the structure of
\begin{align}
\label{eq:04}
\hat{\mathbf{v}}_k=\sqrt{p_k}\frac{(\mathbf{I}_N+\sum_{k=1}^K\frac{\lambda_k}{\sigma^2}\mathbf{h}_k\mathbf{h}_k^H)^{-1}\mathbf{h}_k}{||(\mathbf{I}_N+\sum_{k=1}^K\frac{\lambda_k}{\sigma^2}\mathbf{h}_k\mathbf{h}_k^H)^{-1}\mathbf{h}_k||_2},\forall k,
\end{align} 
where $\boldsymbol{\lambda} = [\lambda_{1},\ldots,\lambda_{K}]^{T}$ is a virtual power allocation vector satisfying that $\sum_{k=1}^{K} \lambda_{k} = \sum_{k=1}^{K} p_{k} = P$. However, such a solution, which estimates the optimal vector $\hat{\mathbf{v}}_{k}$ at BS by inputting the optimum $\{\boldsymbol{\lambda},\mathbf{p}\}$ predicted by using DNN-based solutions \cite{9926157,10146432} into (\ref{eq:04}) will incur prohibitively high computational cost of $\mathcal{O}(N^{3})$, particularly for the large antenna array $N$ in a given scene.

\begin{figure}[tp]
 \setlength{\abovecaptionskip}{-0.1cm}
\setlength{\belowcaptionskip}{-0.4cm}
    \begin{center}
    \includegraphics[width=9.5cm]{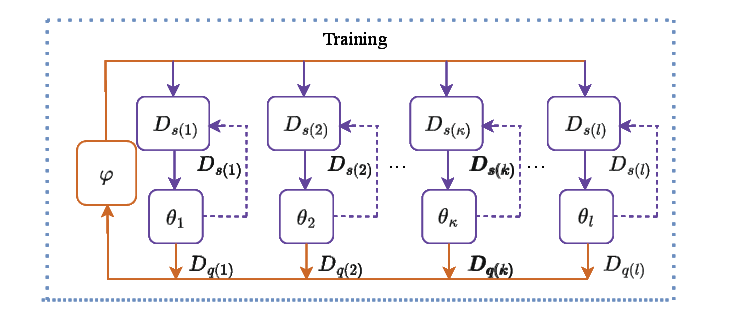}
    \end{center}
 \caption{An overview of model-agnostic meta-learning framework.}\label{fig:002}
\end{figure}

\subsection{Meta Learning-Inspired Beam Prediction}

As illustrated in Fig. \ref{fig:002}, the MAML framework in traditional ML-aided beam prediction starts by defining a set of tasks, each involving prediction from a set of channel realizations \cite{Finn2017ModelAgnosticMF}. Each task has corresponding training data (i.e., the support set denoted by $\mathcal{D}_{s}$) and validation data (i.e, the query set denoted by $\mathcal{D}_{q}$). MAML then uses \textit{two optimization loops} to obtain a meta-initialization for dynamic adaptation on new tasks. 

The \textit{inner loop} is used to update the parameters of each task based on the relevant support set $\mathcal{D}_{s}$. 
% while the \textit{outer loop} is used to update the global neural network parameters based on the query sets $\mathcal{D}_{q}$ of all tasks. 
Let $Loss_{\mathcal{D}_{s}(\kappa)}\left(\theta_{\kappa}\right)$ be the performance metric for task $\kappa$ in the inner loop.
Then, each task optimizes its parameters independently, typically using GD technique to minimize the loss function on its support set $\mathcal{D}_{s}$ \cite{Finn2017ModelAgnosticMF}, i.e.,
\begin{align}
\label{eq:005}
\hat{\theta}_{\kappa}=\mathop{\arg\min}\limits_{\theta_\kappa} ~ Loss_{\mathcal{D}_{s}(\kappa)}\left(\theta_{\kappa}\right),\quad\forall \kappa.
\end{align}  

In the \textit{outer loop}, MAML updates the global network parameters with the goal of minimizing the sum of the loss functions of all tasks on their query set $\mathcal{D}_{q}$. Likewise, let $Loss_{\mathcal{D}_{q}}(\psi)$ be the performance metric for the tasks in the outer loop.
Once completing training, it enters the meta-testing phase for adaptation. 
In this phase, the pre-trained global network parameters are used to adapt to the new task.
Note that the process of updating the parameters of the new task is similar to an inner loop, but using the adaptation dataset. 
The parameters of the new task are updated based on the adaptation dataset and optimized via GD technique \cite{Finn2017ModelAgnosticMF}, i.e.,
\begin{align}
\label{eq:006}
\hat{\psi}=\mathop{\arg\min}\limits_{\psi} ~ Loss_{\mathcal{D}_{q}}(\psi).
\end{align}

Note that the core task of the ML algorithm in the outer loop phase is to update, through an iterative process, the global network parameters $\varphi$, which serve as the initial point for all tasks, with the goal of finding a parameter configuration that adapts quickly on different tasks. 
% This process focuses on optimizing the generalization ability of the model, i.e., its adaptability on new tasks. 
By calculating the average loss of the updated parameters in the inner loop over the query set of each task, the outer loop iteratively updates $\varphi$ to minimize this loss. This process does not involve updating the parameters individually for each specific task, but rather extracts generic features and knowledge from the meta-training dataset so that the model can achieve effective adaptation on new tasks with a minimal amount of data.

% Furthermore, the key to the outer loop is to capture commonalities across tasks and utilize these commonalities to guide the rapid adaptation of the model. This approach is particularly suitable for tasks in dynamic environments, where models need to frequently adapt to new situations. In this way, ML algorithm as well as its variants are able to learn how to extract useful features and knowledge from previous tasks and migrate this knowledge to new tasks, thus enabling fast learning and adaptation on new tasks.

By conducting the above two optimization loops, the optimal beam associated with the underlaying channel distribution then can be determined with low complexity. However, one of the main drawbacks of the MAML algorithm is that it is complicated in the training and testing phases. Furthermore, when encountering divergent training and testing channel distributions, MAML's adaptation process will become to be slow, thus severely degrading the accuracy performance.

%------------------------------> Fig.3
\begin{figure*}[t!]
\setlength{\abovecaptionskip}{-0.1cm}
\setlength{\belowcaptionskip}{-0.4cm}
    \begin{center}
    \includegraphics[width=16.5cm,height=7cm]{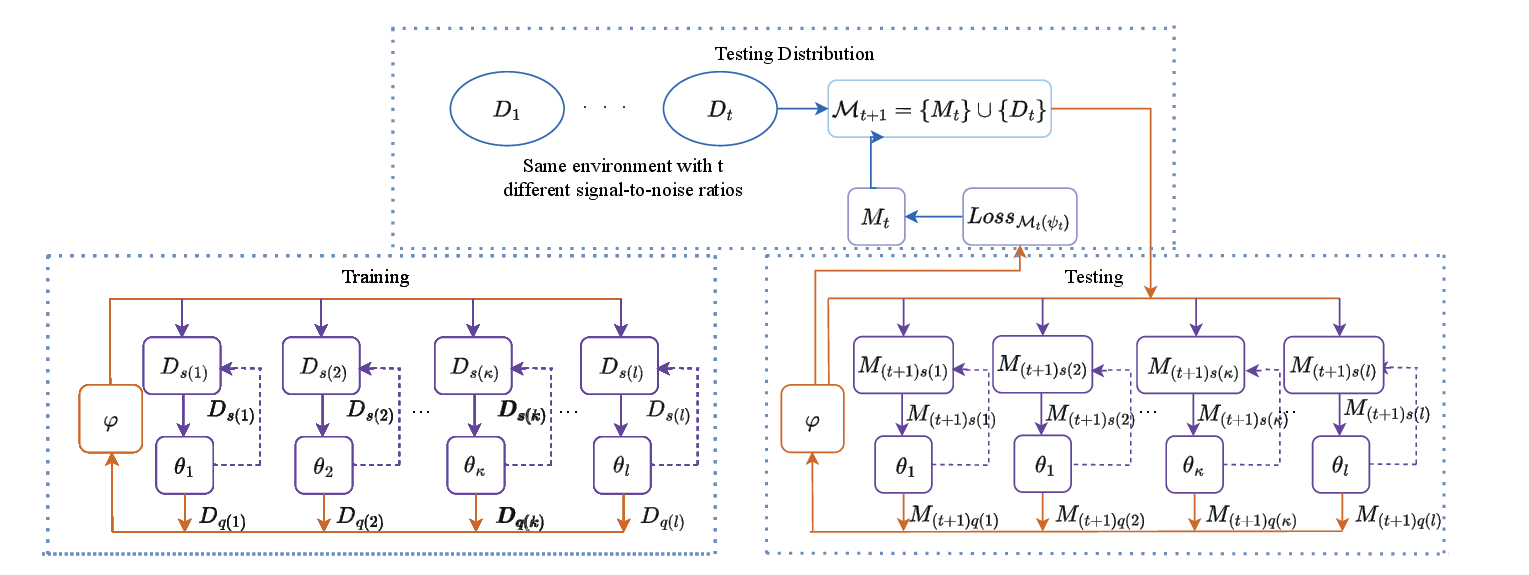}
    \end{center}
   \caption{An overview of the training and testing phases of the M-ML framework.}\label{fig:03}
\end{figure*}
%--------------------------------

\subsection{The Design Rational of Our Proposal}

When using a small amount of data, traditional ML based solutions are prone to overfitting. Moreover, since the new environment encountered may be quite different from the training environment when considering the small-scale fading, it is often difficult for the model to achieve optimum performance at the initial stage, and thus it takes longer time to adapt to a new scenario. In this paper, we propose an innovative memory set selection mechanism to update the data, specifically, selecting poorly performing data as a selected representative sample through a loss function defined in (\ref{eq:012}). Usually, these data points  contain information that the model has not yet adapted to or understood in the new environment.
Accordingly, the latest test data and representative samples are combined to form a memory set $\mathcal{M}_{t}$, whose size is $M$. In this sense, the  memory set is dynamically updated at time instance $t+1$ according to
\begin{align}
\label{eq:0007}
    \mathcal{M}_{t+1}=\left\{D_{t}^{\left(i\right)}\right\}_{i\in\Omega_{Dt}}\cup\left\{M_{t}^{\left(i\right)}\right\}_{i\in\Omega_{Mt}},
\end{align}
where $D_{t}^{(i)}$ denotes the test data collected at time instance $t$, having the characteristics and patterns of
the new environment. 
$\Omega_{Dt}$ and $\Omega_{Mt}$ mean the index sets of the latest test data $\mathcal{D}_{t}$ and selected representative samples $\mathcal{M}_{t}$, respectively. 
The selection of data ensures that the selected samples capture key features of the environment. Then, the proposed M-ML framework updates the global network parameter in the testing phase by minimizing the sum of the loss functions
of all tasks on their updated query set $\mathcal{M}_{t}$ at the time instance $t$ under the total memory size constraint:\footnote{
Note that in the proposed M-ML framework, memristors play a critical role in i) enhancing computational efficiency, ii) reducing latency, and iii) enabling dynamic adaptation in beam prediction for non-stationary mmWave communications. Specifically, memristors enable online learning by allowing on-the-fly updates to stored memory samples $\mathcal{M}_{t}$, ensuring the model retains knowledge of recent channel conditions. This contrasts with traditional ML approaches that require full retraining when environment changes.
}
\begin{align}
\label{eq:0008}
\begin{split}
&\mathop{\text{minimize}}_{\psi_{t}}\quad Loss_{\mathcal{M}_{t}}(\psi_{t})\\
&s.t. \; \left|\left\{{D}_{t}^{\left(i\right)}\right\}_{i\in\Omega_{Dt}}\cup\left\{{M}_{t}^{\left(i\right)}\right\}_{i\in\Omega_{Mt}}\right| \leq M, \;\; t=1,2,\ldots
\end{split}
\end{align}

Based on the above discussions and following the MAML-aided beam prediction approaches \cite{10096315,9257198}, we attempt to solve the joint beam prediction and the memory set selection problem  by following three consecutive stages: i) beamforming matrix decomposition, ii) M-ML model training and testing phases, iii) beamformer reconstruction. Accordingly, the joint beam prediction and the memory set selection problem is divided into the following cascaded subproblems:
\begin{itemize}
\item \textbf{Beamforming matrix decomposition:} We decompose the beamforming matrix $\mathbf{V}$ of size $N\times K$ into three low-dimensional components $\mathbf{w}$ of size $K\times $1 in (\ref{eq:0009}), $\mathbf{u}$ of size $K\times $1 in (\ref{eq:0010}), and $\mu$.   
\item \textbf{M-ML model training and testing:} M-ML predicts the low-dimensional components $\mathbf{w}_{}$,  $\mathbf{u}_{}$, and $\mu$ by training the model parameters $\theta_{u}^{l}$ in (\ref{eq:0013}), $\theta_{w}^{l}$ in (\ref{eq:0014}), and $\theta_{\mu}^{l}$ in (\ref{eq:0015}) using an updated dataset $\mathcal{M}_{t}$.   
\item \textbf{Beamformer reconstruction:} We reconstruct the beamforming matrix using these predicted components according to (\ref{eq:009}).
\end{itemize}

% The above joint beamforming and memory set selection optimization is summarized in Algorithm 1. 
More details regarding the joint beamforming and memory set selection optimization   are explained in the following Section III. 

% \alglanguage{pseudocode}
% \begin{algorithm}[!t]\small
% \caption{Summary of Joint Beam Prediction and Memory Set Selection}
% \label{alg:Framwork2}
% \begin{algorithmic}[1]
% % \Require
% % Initialize the memory set $M_0$ = $\varnothing$, memory set size $M$, counter of time slots $t$ = 0, the index sets of the selected samples $\Omega_{D_t}$, and the index sets of the selected samples $\Omega_{M_t}$
% % \Ensure The non-stationary file popularity $\hat{\mu}^{}_{1:F,1:T}$.
% \For{{$t$} = 1, 2, $\ldots$}
% \State \multiline{Decompose
% the beamforming matrix $\mathbf{V}_t$ into three low-dimensional components based on the WMMSE algorithm.}
% \State \multiline{Learn the decomposed low-dimensional components by using the updated memory set $\mathcal{M}_{t}$.}
% \State \multiline{Reconstruct the beamforming matrix $\hat{\mathbf{V}}_t$ towards the new channel distribution using the predicted results in step 3.}
% \EndFor
% \end{algorithmic}
% \end{algorithm}
% %-------------------------------->

% \begin{figure}[tp]
% % \setlength{\abovecaptionskip}{-0.5cm}
% % \setlength{\belowcaptionskip}{-0.2cm}
%     \begin{center}
%     \includegraphics[width=8cm]{fig/re3.pdf}
%     \end{center}
%    \caption{An overview of the meta-regularization framework.}\label{fig:02}
% \end{figure}

\section{M-ML Aided Beam Prediction}
In this Section, we propose an innovative M-ML framework to expedite spatial and temporal domain beam prediction, as illustrated in Fig. \ref{fig:03}. The main objective of this framework is to combine the generalization ability of ML pre-trained models with the storage property of memristors. 
When deploying the pre-trained model into a new test environment, our approach utilizes the memory to dynamically store and process data in the new environment. 
In this case, our approach can not only maintain high performance on the original task, but also show good generalization ability and robustness in the new test environment.\footnote{Note that Bayesian meta-learning (BML) is a framework that integrates Bayesian principles into ML to address uncertainty and improve generalization in low-resource tasks \cite{yap2021addressing}. As a major difference from M-ML, BML takes into account i) the latest dataset $\mathcal{D}_t$ from the new environment, and ii) the posterior of the meta hypeparameters during optimization. We will investigate the BML-based beam prediction technique as well as its scalability and robustness in realistic mmWave massive MIMO in our future research.} 
More details are stated as follows.

\subsection{Beamforming Matrix Decomposition}
Along with an increasing number of transmitting antennas and users, estimating the beamforming vectors directly at BS by using DNN-based solutions \cite{9926157,10146432} aggravates the training burden of the neural networks. To tackle this challenge, we attempt to decompose the beamforming matrix  $\mathbf{V} = \left[\mathbf{v}_{1},\ldots,\mathbf{v}_{K}\right]$ into three low-dimensional components, i.e., $w_{k}$, $u_{k}$, and $\mu_{}$. 
The expressions of $w_k$ and $u_k$ are as follows:
\begin{equation}
\label{eq:0009}
w_k=
\frac{\sigma^2+{\sum_{j=1}^N\left|\mathbf{h}_k^H\mathbf{v}_j\right|}^{2}}{{\sigma}^{2} + \sum_{j=1,j\neq k}^N{\left|{\mathbf{h}_k^H}\mathbf{v}_j\right|}^{2}},
\end{equation}
\begin{equation}
\label{eq:0010}
u_k=
\frac{\mathbf{h}_k^H\mathbf{v}_k}{{\sigma}^{2} + \sum_{j=1}^N{\left|{\mathbf{h}_k^H}\mathbf{v}_j\right|}^{2}}.
\end{equation}
While $\mu$ $\geq$ 0 is a Lagrange multiplier that exists in the power constraints when finding the first-order optimality condition of the beamforming matrix in the WMMSE algorithm \cite{5756489}. 

In the training phase, we employ three independent neural networks, each of which is responsible for predicting three basic components of the beamforming vectors, i.e.,  $w_{k}$, ${u}_k$, and $\mu$. The training starts with a random initial value of $\mathbf{V}$, followed by the predictions of $w_k$ and $u_k$ using the current estimate of $\mathbf{V}$. %while $\mu$ is determined by a dichotomous search method, which is different from the neural network prediction approach employed in our method. 
Once finishing the above prediction process, these components are used to reconstruct the beamforming vector $\hat{\bar{\mathbf{v}}}_{k}$ of UE $k$ by means of  
\begin{equation}
\label{eq:009}
\hat{\bar{\mathbf{v}}}_k=\alpha_ku_kw_k{\mathbf{h}_k}\left(\mathbf{S}+\mu\mathbf{I}_{N}\right)^{-1},
\end{equation}
where $\mathbf{S} := \sum_{k=1}^K\alpha_k|u_k|^2w_k\mathbf{h}_k\mathbf{h}_k^H$. Accordingly, the power constraint is satisfied by normalizing the transmit beamforming matrix $\hat{\bar{\mathbf{V}}} = [\hat{\bar{\mathbf{v}}}_1,\ldots,\hat{\bar{\mathbf{v}}}_K]$ and then iteratively updating it according to the above equations.

It seems that the above updating process is similar to the WMMSE algorithm \cite{5756489}, which continues until a preset error criterion is reached. However, the difference is that: i) our approach simplifies the iterative process by directly predicting $u_k$, $w_k$, and $\mu$ through neural networks and then reconstructs the value of $\hat{\bar{\mathbf{V}}}$ based on these predictions.
ii) In the training phase, we specifically design the loss function, which is defined as follows: 
% \begin{equation}
% \label{eq:012}
% \begin{split}
% Loss(v)=&-\sum_{i=1}^{K}\text{log}~\text{det}\Big(\mathbf{I}+\\
% &\mathbf{h}_i\mathbf{v}_i\mathbf{v}_i^H\mathbf{h}_i^H\Big(\sum_{j\neq i}^{K}\mathbf{h}_j\mathbf{v}_j\mathbf{v}_j^H\mathbf{h}_j^H+\sigma^2\mathbf{I}\Big)^{-1}\Big),
% \end{split}
% \end{equation}
\begin{equation}
\label{eq:012}
\begin{split}
Loss(\mathbf{V})=&-\frac{1}{K}\sum\nolimits_{i=1}^{K}\text{log}~\Big(1+\\
&\mathbf{h}^{H}_i\mathbf{v}^{}_i\mathbf{v}^{H}_i\mathbf{h}_i\Big(\sigma^2 + \sum_{j\neq i}^{K}\mathbf{h}^{H}_j\mathbf{v}_j\mathbf{v}_j^H\mathbf{h}_j\Big)^{-1}\Big).
\end{split}
\end{equation}

\subsection{Inner Model Training}
In each training cycle, we randomly select $N_s$ and $N_q$ data points from the channel realizations of each task to construct the support set $\mathcal{D}^i_s$ and the query set $\mathcal{D}^i_q$, respectively, where $i$ indicates the index of the channel groups. For each single training epoch, the parameters  $\theta_u$, $\theta_w$, and $\theta_{\mu}$ experience total numbers of $N_t$ updates.
Likewise, define $\theta_u$, $\theta_w$, and $\theta_{\mu}$ as the parameters related to the inner models trained by a total number of $N_{t} \times (N_s + N_q)$ channel realization samples as entry to predict $u_{k}$, $w_k$, and $\mu$, respectively, with $k\in\{1,\ldots,K\}$. Define $\varphi_u$,  $\varphi_w$, and  $\varphi_{\mu}$ as the parameters related to the outer models.

In the inner loop, we process a total number of $N_t \times (N_s + N_q)$ channel realization samples as input to predict $u_{k}$, $w_k$, and $\mu$ for $\forall k\in\{1,\ldots,K\}$. 
We then reconstruct the beamforming vectors and normalize them to satisfy the total power requirement in (\ref{eq:003}), and finally compute the sum rate as well as the loss for each user.
The inner model parameters are updated following the criteria of 
\begin{equation}
\label{eq:0013}
\theta_{u}^{i}=\theta_{u}^{i-1}-a\nabla_{\theta_{u}}\sum\nolimits_{d=1}^{N_{s}}Loss(\mathbf{V}_{d}^{i}),
\end{equation}
\begin{equation}
\label{eq:0014}
\theta_{w}^{i}=\theta_{w}^{i-1}-a\nabla_{\theta_{w}}\sum\nolimits_{d=1}^{N_{s}}Loss(\mathbf{V}_{d}^{i}),
\end{equation}
\begin{equation}
\label{eq:0015}
\theta_{\mu}^{i}=\theta_{\mu}^{i-1}-a\nabla_{\theta_{\mu}}\sum\nolimits_{d=1}^{N_{s}}Loss(\mathbf{V}_{d}^{i}),
\end{equation}
where $a$ is the learning rate of the inner model and $\mathbf{V}_d^i$ represents the beamforming matrix realized for the $d$th channel in the $i$th set. At the beginning of each period, the parameter updates are influenced by the last set of data from the previous period. The parameters that undergo updates subsequent to processing the last set of data in the current period are considered as the definitive parameters for that particular period.

\subsection{Outer Model Testing}
Once finishing the inner loop, the model enters the outer loop, and 
the corresponding outer model parameters are updated following the criteria of \footnote{
As a practical communication scenario which is episodically dynamic is considered in our MISO-BC systems, we add a superscript "e" in the outer model parameters.   
}
\begin{equation}
\label{eq:016}
\varphi_u^e=\theta_u^e-\beta\nabla_{\theta_u}\sum\nolimits_{i=1}^{N_t}\sum\nolimits_{d=1}^{N_q}Loss(\mathbf{V}_d^{e,i}),
\end{equation}
\begin{equation}
\label{eq:017}
\varphi_{w}^{e}=\theta_{w}^{e}-\beta\nabla_{\theta_{w}}\sum\nolimits_{i=1}^{N_{t}}\sum\nolimits_{d=1}^{N_{q}}Loss(\mathbf{V}_{d}^{e,i}),
\end{equation}
\begin{equation}
\label{eq:018}
\varphi_{\mu}^{e}=\theta_{\mu}^{e}-\beta\nabla_{\theta_{\mu}}\sum\nolimits_{i=1}^{N_{t}}\sum\nolimits_{d=1}^{N_{q}}Loss(\mathbf{V}_{d}^{e,i}),
\end{equation}
where the parameter $\beta$ is the learning rate of the outer model, and $\mathbf{V}_d^{e,i}$ accounts for the beamforming matrix corresponding to the $d$th data point of group $i$ at the $e$th epoch.

During the testing period, we dynamically add the test data $D_{t}^{(i)}$ collected in real time to the memory set $\mathcal{M}^{(i)}_{t}$.\footnote{These new data capture the important characteristics and subtle patterns of the new environment.} Subsequently, we combine these new data with the existing data into $\mathcal{M}^{(i)}_{t}$ which serves as the input to the neural networks to jointly participate in the next round of testing and training. This approach not only enriches the model's knowledge of the new environment, but also provides more information for the model to adjust and optimize its own parameters,
thus achieving more accurate prediction and higher performance in the new environment.

\subsubsection{\textbf{Loss Sensitivity-Based Memory Updating Mechanism}}
To further improve the model's adaptability in the new environment, we adopt a loss-based ranking strategy. According to (\ref{eq:012}), we evaluate and rank the losses incurred by the model during testing, followed by identifying the data points with the most severe losses by solving the optimization problem of 
\begin{align}
\label{eq:019}
\hat{\Omega}_{Mt} = \mathop{\arg\max}\limits_{\mathcal{M}_{t}} ~ Loss_{}(\mathbf{V}_{d}^{e,i}).
\end{align}
These data points usually contain information that the model has not yet adapted to or understood in the new environment, and thus have a high learning value. We prioritize these data points with the highest losses in memory set $\mathcal{M}_{t}$ to ensure that the model can focus on these difficult samples during the next round of training. Given that the BS's storage capacity (i.e., the memory size) is $M$, then the top $M-\pi(t)$ highest ranking channel groups are selected and added to the memory set $\mathcal{M}_{t}$, where $\pi(t)$ accounts for the samples selected from $D_{t}^{(i)}$.   
In this way, the model is not only able to quickly learn key features in new environments, but also gradually reduces its sensitivity to false predictions. This memory update strategy based on loss sensitivity effectively improves the model's adaptability and accuracy in the face of new and unknown environments, providing strong support for the robustness of ML algorithms in practical applications. Note that the above procedure is summarized and detailed in the \textbf{Algorithm \ref{alg:Framwork2}}.\footnote{
Note that the full iterative WMMSE algorithm \cite{5756489} will involve a complexity cost of  $\mathcal{O}(n\cdot N^3)$, where $n$ denotes the number of iterations of the algorithm, thus rendering the estimated beamforming weights outdated as the small-scale fading varies in the order of milliseconds. The DNN-based solution \cite{9926157} will incur prohibitively high computational cost of $\mathcal{O} (N^3)$. Besides, the model-based deep unfolding method will incur a complexity order of $\mathcal{O} (nN_L N^2)$, with $N_L$ being the number of inner iterations in each iteration of the algorithm. However, the proposed approach can accelerate the convergence by reducing the dimension of the predicted variables from $2(N × K)$ to $2 K$. Hence, we deduce that the proposed approach will achieve superior convergence performance than WMMSE \cite{5756489}, DNN-based solution \cite{9926157}, and the model-based deep unfolding method.}

\alglanguage{pseudocode}
\begin{algorithm}[!t]
\caption{\small Proposed M-ML Aided Beam Prediction Framework}
\label{alg:Framwork2}
\begin{algorithmic}[1]
\small
\Require
Initialize the memory set $\mathcal{M}_0$ = $\varnothing$, memory set size $M$, beamforming matrix $\mathbf{V}$, counter of time slots $t$ = 0, the index sets of the selected samples $\Omega$.
\Ensure The index set $\hat{\bar{\Omega}}^{}_{M_{t}}$ and the recovered beamformer $\hat{\bar{\mathbf{V}}}_{}$.

\State
$\;\vartriangleright$ \textbf{Beamforming matrix decomposition}
\State \multiline{Decompose the beamforming matrix $\mathbf{V}$ into low-dimensional components $\mathbf{w}$, $\mathbf{u}$, and $\mu$. }
$\vartriangleright$ \textbf{Low-dimensional component prediction}
\For{{$T$} = 1, 2, $\ldots$}
\State \multiline{Select $N_s$ and $N_q$ data from the channel realizations of each task at random to construct the support set $\mathcal{D}^i_s$ and the query set $\mathcal{D}^{i}_{q}$, respectively, to form a memory set $\mathcal{M}_T \leftarrow  \mathcal{D}^i_s \cup \mathcal{D}^{i}_{q}$. 
 }
\For{$t = 1, \ldots, T$}
\State\multiline{Predict the instantaneous values of low-dimensional components $\mathbf{w}_{t}$ in (\ref{eq:0009}), $\mathbf{u}_{t}$ in (\ref{eq:0010}), and $\mu_{t}$.}
 \State \multiline{Reconstruct the beamforming matrix $\hat{\bar{\mathbf{V}}}_{t}$ based on these predicted low-dimensional components by (\ref{eq:009}).}
 \State \multiline{Update the inner model parameters $\theta^{t}_{u}$, $\theta^{t}_{w}$, and $\theta^{t}_{\mu}$ by minimizing the loss function $Loss(\hat{\bar{\mathbf{V}}}_{t})$ in (\ref{eq:012}).}
\EndFor
\For{$t = 1, \ldots, T$} 
\State \multiline{Predict the instantaneous value of low-dimensional components $\mathbf{w}_{t}$, $\mathbf{u}_{t}$, $\mu_{t}$, and reconstruct the beamforming matrix $\hat{\bar{\mathbf{V}}}_{t}$ based on these predicted low-dimensional components.}
\State \multiline{Update the outer model parameters $\varphi_u^e$, $\varphi_w^e$, and $\varphi_{\mu}^e$ by minimizing the loss function $Loss(\hat{\bar{\mathbf{V}}}_{t})$ in (\ref{eq:012}).}
\EndFor
\State \multiline{Select $\pi(t)$ samples from $D_{t}^{(i)}$ and then update the index set of the selected sample $\Omega_{D_{t}}$.}
\State \multiline{Identify the data points with the most severe
losses by searching the index
sets $\hat{\bar{\Omega}}_{M_{t}}$ according to
\begin{align}
    \small
\hat{\bar{\Omega}}_{M_{t}}
    \leftarrow
    \textrm{Rank}_{M-\pi\left(t\right)}\left(Loss(\hat{\bar{\mathbf{V}}}_{t})\right).
\end{align}
}
\State \multiline{Update the memory set $ \mathcal{M}_{T+1}\leftarrow\{D_{t}^{\left(i\right)}\}_{i\in\Omega_{Dt}}\cup\{M_{t}^{\left(i\right)}\}_{i\in\hat{\bar{\Omega}}_{Mt}}.
$}
\State \multiline{Normalize
the recovered beamformer to satisfy the total power requirement in (\ref{eq:003}) by
\begin{align}
\small
\label{equ:04091421}
\hat{\bar{\mathbf{V}}}_{t} \leftarrow \frac{\hat{\bar{\mathbf{V}}}_t}{\|\hat{\bar{\mathbf{V}}}_t\|_{2}} \sqrt{P}.
\end{align}
}
\EndFor
\end{algorithmic}
\end{algorithm}
%-------------------------------->

% %-------------------------------->
\subsubsection{\textbf{$k$-NN Based Memory Updating Mechanism}}
Alternatively, we employ a multiclass classification algorithm, i.e., $k$-NN \cite{friedman1977algorithm}, to classify the channel distributions from %the $N_{t}\times (N_{s}+N_{q})$ channel realization samples 
% in given scenario 
the query record set
into $M_{d}$ categories and then rank them in order. Afterward, we select the top $M-\dot{\pi}(t)$ highest ranking channel groups as the candidate data points for outer model testing, followed by adding these candidate data points into the memory set $\hat{\bar{\mathcal{M}}}_t$. Herein,
$\dot{\pi}(t)$ represents the samples selected from the test dataset.
The query record can similarly be represented as a point $\mathbf{X}_q$ in this space.
Our objective is to dynamically search the optimal channel distribution (i.e., the $M_d$ categories) having the largest Euclidean distance away from the newly added test data, 
denoted by $\mathbf{Y}_q$ from 
$\mathcal{D}_q$. 
%of the channel distribution $i$ in a new scene.
Notably, the proposed $k$-NN based memory updating strategy enforces that the predicted mmWave beamforming vectors
are well-suited to episodically dynamic channel distribution for the following reasons: i) the proposed memory update strategy enhances the diversity and representativeness of the data samples collected in the memory set $\hat{\bar{\mathcal{M}}}_t$, which in return helps model in learning more robust and generalizable features;  
ii) this enables the model to adequately capture the various situations that may arise in practical applications, thus leading to an improvement in prediction accuracy.\footnote{
Empirical evidence in \cite{friedman1977algorithm} shows that this classification technique is considerably faster than other methods. The total complexity involved in the testing phase of M-ML associated with $k$-NN based memory updating strategy is $\mathcal{O}(N_m^2)$, with $N_m$ being the number of samples in the query record set $\mathcal{D}_q$. However, advanced multi-class classification techniques, such as naive Bayesian model (NBM), assume that features are independent of each other, which is often difficult to meet in practical applications.
}

To this end, we formulate the channel distribution matching problem as follows.
The best match problem is then to find the $m$ points to the query point in this vector space of $l_p$-norms by using the following dissimilarity measures: 
% \begin{align}
% \label{eq:022}
%     D_{p}(\mathbf{X}_q, \mathbf{Y}_q) =\left[\sum_{i=1}^{k}\mid \mathbf{X}_q(i) - \mathbf{Y}_q(i)\mid^{p}\right]^{\frac{1}{p}}.
% \end{align}
\begin{align}
\label{eq:022}
    D_{p}(\mathbf{X}_q, \mathbf{Y}_q) =\left[\sum_{i=1}^{k}\left\| \mathbf{X}_q(i) - \mathbf{Y}_q(i)\right\|_{p}\right]^{\frac{1}{p}}.
\end{align}
% We observe that, the performance of the proposed algorithm may depend upon the total number of records in the file $N$, the dimensionality (number of keys) $k$, the number of the nearest neighbors sought $m$, the number of records in the terminal buckets $\mathbf{b}$, the dissimilarity measure $D(\mathbf{X},\mathbf{Y})$ employed, and the distribution $p(\mathbf{X})$ of the file records in the record key space.
Herein, it is noted that when $p$ = 2, the dissimilarity measures $D_2(\mathbf{X}_q, \mathbf{Y}_q)$ accounts for the
mostly commonly used Euclidean distance.\footnote{Notably, when $p$ = 2, this dissimilarity measures $D_2(\mathbf{X}_q, \mathbf{Y}_q)$ can also be viewed as the $l_2$-norm of the matrix $\mathbf{X}_q(i) - \mathbf{Y}_q(i)$, $i\in\{1, \ldots,k\}$.} When taking the special case of $p =\infty$ into accounts, the dissimilarity measures $ D_{p}(\mathbf{X}_q, \mathbf{Y}_q)$ in (\ref{eq:022}) will become to be the maximum coordinate distance, i.e., $D_{\infty}(\mathbf{X}, \mathbf{Y}) = \max_{1\leq i\leq k}\left\| \mathbf{X}(i) - \mathbf{Y}(i)\right\|$, which results in a lower bound for all vector space $l_p$-norms \cite{friedman1977algorithm}.
Accordingly, the candidate data point having the largest Euclidean distances, i.e., $p=2$, from the newly added data in new channel environment can be selected by solving the optimization problem of  
\begin{align}
\begin{split}
\label{eq:023}
\hat{\Phi}_{M_t} =& \mathop{\arg\min}\limits_{\hat{\bar{\mathcal{M}}}_{t}} ~-D_{2}(\mathbf{X}_q, \mathbf{Y}_q)\\
=& \mathop{\arg\max}\limits_{\hat{\bar{\mathcal{M}}}_{t}} ~ \left[\sum_{i=1}^{k}\left\| \mathbf{X}_q(i) - \mathbf{Y}_q(i)\right\|_{2}\right]^{\frac{1}{2}},
\end{split}
\end{align}
where $\hat{\Phi}_{M_t}$ indicates the index set of selected candidate representative samples $\mathcal{M}_t$ at the time instance $t$.

% We observe that, the performance of the proposed algorithm may depend upon the total number of records in the file $N$, the dimensionality (number of keys) $k$, the number of the nearest neighbors sought $m$, the number of records in the terminal buckets $\mathbf{b}$, the dissimilarity measure $D(\mathbf{X},\mathbf{Y})$ employed, and the distribution $p(\mathbf{X})$ of the file records in the record key space.

\alglanguage{pseudocode}
\begin{algorithm}[!t]
\caption{\small Proposed $k$-NN Based Memory Updating Approach}
\label{alg:Framwork3}
\begin{algorithmic}[1]
\small
\Require
Initialize the memory set $\mathcal{M}_0$ = $\varnothing$, memory set size $M$, beamforming matrix $\mathbf{V}$, counter of time slots $t$ = 0, the index sets of the selected samples $\Phi$.
\Ensure The index set $\hat{\Phi}^{}_{M_{t}}$ and the recovered beamformer $\hat{\bar{\mathbf{V}}}_{}$.
\State \multiline{Decompose the beamforming matrix $\mathbf{V}$ into low-dimensional components $\dot{\mathbf{w}}$, $\dot{\mathbf{u}}$, and $\dot{\mu}$. }
% $\vartriangleright$ \textbf{Low-dimensional component prediction}
\For{{$T$} = 1, 2, $\ldots$}
\State \multiline{Execute the step 4 in \textbf{Algorithm 1} to construct a memory set $\hat{\bar{\mathcal{M}}}_T \leftarrow  \mathcal{D}^i_s \cup \mathcal{D}^{i}_{q}$. 
 }
\For{$t = 1, \ldots, T$}
 \State \multiline{Update the inner model parameters $\theta^{t}_{\dot{u}}$, $\theta^{t}_{\dot{w}}$, and $\theta^{t}_{\dot{\mu}}$ by conducting steps 6-8 in \textbf{Algorithm 1}.}
\EndFor
\For{$t = 1, \ldots, T$} 
\State \multiline{Update the outer model parameters $\varphi_{\dot{u}}^e$, $\varphi_{\dot{w}}^e$, and $\varphi_{\dot{\mu}}^e$ by conducting steps 11-12 in \textbf{Algorithm 1}.}
\EndFor
\State \multiline{Select $\dot{\pi}(t)$ samples from $D_{t}^{(i)}$ and then update the index set of the selected sample $\Omega_{D_{t}}$.}
\State \multiline{Identify the data points with the largest dissimilarity measurements by searching the index
sets $\hat{\bar{\Phi}}_{M_{t}}$ according to
\begin{align}
    \small
\hat{\bar{\Phi}}_{M_{t}}
    \leftarrow
    \textrm{Rank}_{M-\dot{\pi}\left(t\right)}\left(\mathop{\arg\min}\limits_{\hat{\bar{\mathcal{M}}}_{t}} ~-D_{2}(\mathbf{X}_q, \mathbf{Y}_q)\right).
\end{align}
}
\State \multiline{Update the memory set $ \hat{\bar{\mathcal{M}}}_{T+1}\leftarrow\{D_{t}^{\left(i\right)}\}_{i\in\Omega_{Dt}}\cup\{M_{t}^{\left(i\right)}\}_{i\in\hat{\bar{\Phi}}_{Mt}}.
$}
\State \multiline{Normalize
the recovered beamformer to satisfy the total power requirement in (\ref{eq:003}) by using (\ref{equ:04091421}).
% $
% \small
% \label{equ:040914}
% \hat{\bar{\mathbf{V}}}_{t} \leftarrow \frac{\hat{\bar{\mathbf{V}}}_t}{\|\hat{\bar{\mathbf{V}}}_t\|_{2}} \sqrt{P}.
% $
}
\EndFor
\end{algorithmic}
\end{algorithm}
%-------------------------------->

Note that the above proposed $k$-NN based memory updating mechanism effectively improves the model's adaptability and accuracy when faced with new and unknown environments, thus capable of providing strong support for the robustness of M-ML algorithms in practical applications. Note also that the above procedure is summarized and detailed in the \textbf{Algorithm \ref{alg:Framwork3}}.

% \subsection{Computational Complexity Analysis}
% %
% It is noted that predicting the high dimensional beamforming matrix $\mathbf{V}$ of size $N \times K$ directly using DNN-based solutions \cite{9926157,10146432} will incur high training overhead particularly when the network size becomes large \cite{8935405}. To reduce the training complexity involved, we predict low-dimensional components of the beamforming vectors using the proposed M-ML framework with low training overhead instead of predicting beamforming matrix $\mathbf{V}$. 
% We note that the dimension of the predicted variables is reduced from $2(N\times K)$ to $2K$. On the other hand, the computational cost incurred for updating the memory is calculated as follows. For the proposed loss sensitivity-based memory updating strategy, the computational cost is incurred mainly by ranking the losses of model during the testing phase in (\ref{eq:012}), thus yielding a computational complexity of $\mathcal{O}(N_n\textrm{log}(N_n))$ with $N_n$ being the number of samples in the memory set $\mathcal{M}^{(i)}_{t}$.
% The $k$-NN based memory updating strategy will incur a computational cost of $\mathcal{O}(N_{m}^{2})$, where $N_{m}$ denotes the number of samples in the query record set $\mathcal{D}_q$.   
%

\section{Meta-Regularized Aided Beam Prediction}
In ML framework, the original parameter $\theta$ is optimized to $\hat{\theta}$ through inner-loop optimization by fine-tuning the initialization parameter using a small number of support set $\mathcal{D}_s$ when encountering a new training task. 
To mitigate meta-overfitting, which occurs when model parameter updates are driven by gradients, one can implement the already existing dropout technique.
% To suppress the meta-overfitting that occurs when the update of model parameters is driven by gradients, one can implement the already existing dropout technique. 
Details are stated as follows.

\subsection{Meta Dropout Regularization}

Note that \textit{dropout regularization} is an approach for preventing overfitting that occurs in DNNs \cite{Srivastava2014DropoutAS} by randomly dropping neurons in the network during training. Specifically, it temporarily removes neurons in DNN with a certain probability, to enhance the generalization ability of the model. The implementation consists of leaving each neuron inactive at each iteration with a hyperparametrically defined probability, while scaling the activation output of the undiscarded neurons in order to maintain network responsiveness. During training, only the retained neurons participate in backpropagation and update the parameters. Dropout method is implemented on the gradients in the inner-loop optimization, denoted as DropGrad, and is not enabled in the outer-loop optimization. All neurons are activated during the testing phase, while the learning rate and the number of iterations may be adjusted to accommodate neuron discarding during the training phase. Note also that the advantages of dropout are that: i) it is simplicity, ii) it is ease of implementation, and iii) it does not need complex parameter tuning, although it may have some impact on training stability and convergence speed.

In the training process of ML, we randomly select a task $\kappa$ from the training set $\mathcal{D}_s$, which is defined by a specific dataset \{$\mathcal{D}_s,\mathcal{D}_q$\}. 
We use a GD algorithm to optimize the initial parameter set $\boldsymbol{\theta}$ of the model to better fit that specific task $\kappa$ \cite{tseng2020regularizing}, i.e.,
\begin{align}
\label{eq:0025}
\hat{\boldsymbol{\theta}}=\boldsymbol{\theta}-\boldsymbol{\alpha}\odot \mathbf{g}.
\end{align}
In this process, the learning rate vector $\boldsymbol{\alpha}$ is used to guide the gradient-based parameter tuning. The gradient vector $\mathbf{g}$ is obtained by evaluating the model's performance on the support set and calculating its gradient w.r.t. the parameters $\boldsymbol{\theta}$, which reflects how well the model fits the training data with the current parameter settings.\footnote{
Reference \cite{tseng2020regularizing} proposes the Gaussian dropout concept, which can augment the gradient with noise sampled from the Gaussian distribution, thus providing a better regularization with uncertainty.
In this work, we integrate the Gaussian dropout concept \cite{tseng2020regularizing} into the model-based meta learning framework to optimize the initial parameter set $\hat{\theta}_{\ddot{u}}$, $\hat{\theta}_{\ddot{w}}$, and $\hat{\theta}_{\ddot{\mu}}$ through (\ref{eq:0025}), as well as to improve the model with high applicability and stability in different environments and conditions.
} 
% In this way, we can ensure that the parameter updates are purposeful, aiming to minimize the discrepancy between the model predictions and the actual data
% \begin{align}
% \mathbf{g}_u=\bigtriangledown_{\varphi_u}L^{s}(f_{\varphi}(\mathbf{X}_{s}),\mathbf{Y}_{s}).
% \end{align}
% \begin{align}
% \mathbf{g}_w=\bigtriangledown_{\varphi_w}L^{s}(f_{\varphi}(\mathbf{X}_{s}),\mathbf{Y}_{s}).
% \end{align}
% \begin{align}
% \mathbf{g}_{\mu}=\bigtriangledown_{\varphi_{\mu}}L^{s}(f_{\varphi}(\mathbf{X}_{s}),\mathbf{Y}_{s}).
% \end{align}

During the training phase, we first perform inner-loop optimization, a gradient-based adaptive process that allows the model to quickly tune its parameters based on the data in the support set $\mathcal{D}_{s}$. Subsequently, we move to outer-loop optimization, where the query set $\mathcal{D}_q$ contains samples not previously seen by the model, providing us with key information to evaluate the model's generalization ability. By calculating the gradient of the model's loss function over the query set $\mathcal{D}_q$, we are able to identify areas where the model needs to improve and use this information to update %the initial parameters
the outer-model parameters by means of the GD method. 
In this way, the outer-model parameters $\varphi_u^e$ in (\ref{eq:016}), $\varphi_w^e$ in (\ref{eq:017}), and $\varphi_{\mu}^e$ in (\ref{eq:018}) can be updated individually at the $e$th epoch, followed by recovering the beamforming matrix $\hat{\bar{\mathbf{V}}}$ accordingly.  
% Thus, we have
% % \begin{align}
% % \boldsymbol{\varphi}=\boldsymbol{\varphi}-\eta\bigtriangledown_{\varphi}L^{q}(f_{\varphi^{\prime}}(\mathbf{X}_{q}),\mathbf{Y}_{q}),
% % \end{align}

% \begin{equation}
% \label{eq:028}
% \varphi_u=\varphi_u-\eta\nabla_{\varphi_u} L^{q}(f_{\varphi^{\prime}}(\mathbf{X}_{q}),\mathbf{Y}_{q}),
% \end{equation}
% \begin{equation}
% \label{eq:029}
% \varphi_{w}=\varphi_{w}^{}-\eta\nabla_{\varphi_{w}}L^{q}(f_{\varphi^{\prime}}(\mathbf{X}_{q}),\mathbf{Y}_{q}),
% \end{equation}
% \begin{equation}
% \label{eq:030}
% \varphi_{\mu}^{}=\varphi_{\mu}^{}-\eta\nabla_{\varphi_{\mu}}L^{q}(f_{\varphi^{\prime}}(\mathbf{X}_{q}),\mathbf{Y}_{q}),
% \end{equation}
% where $\eta$ is the learning rate for meta-training, determining the step size of the parameter update.

\subsection{Meta Gaussian DropGrad}
It is noted that our core strategy in the training phase of ML is to introduce uncertainty in the inner-loop optimization process as a way to enhance the generalization ability of the model. This strategy focuses specifically on the gradient $g$, which plays a key role in inner-loop optimization for tuning the model parameters $\boldsymbol{\theta}$. Our approach is to apply stochasticity to the gradient $g$ during parameter updating so that the updated parameters $\boldsymbol{\hat{\theta}}$ are better able to generalize to the unseen data
\begin{equation}
\label{eq:0026}
\begin{aligned}
\hat{\mathbf{g}}=\mathbf{g}\odot \mathbf{n},
\end{aligned}
\end{equation}
where $\mathbf{n}$ is the noise regularization term sampled from a predetermined distribution. Based on the above equation, we introduce a Gaussian distribution noise regularization strategy below. Taking the expectation and variance of the noise term $\mathbf{n}_g$ in the DropGrad method as $\mathbb{E}(\mathbf{n}_g)=$ l and $\mathrm{Var}(\mathbf{n}_g)=\frac{P_g}{1-P_g}$, respectively, we then add the noise $\mathbf{n}_{g}$ sampled from the Gaussian distribution to
the gradient using the following expression:
\begin{equation}
\label{eq:0027}
\begin{aligned}
\hat{\mathbf{g}}=\mathbf{g}\odot \mathbf{n}_{g},\quad \mathbf{n}_{g}\sim \mathcal{CN}(1,\frac{P_g}{1-P_g}),
\end{aligned}
\end{equation}
where $P_g$ denotes the Gaussian Gradient dropout probability. 

Note that the Gaussian distribution $\mathcal{CN}(1,\frac{P_g}{1-P_g})$, as a means of regularization, can effectively introduce and regulate uncertainty in model training, thus enhancing its generalization ability. In the process of ML training, $\mathbf{n}_{g}$ is introduced through the process of updating the parameters of the model, a strategy that not only increases the robustness of the model, but also simulates the randomness and variability that exists in the real world. In this way, the model is able to adjust more flexibly in the face of new and unseen data, reducing the dependence on specific training data, and thus improving its applicability and stability in different environments and conditions. Note also that the use of Gaussian distributions provides our model with a natural and mathematically rigorous way to quantify and control uncertainty in the training process. The above procedure is summarized and detailed in the \textbf{Algorithm \ref{alg:Framwork4}}.

\subsection{Computational Complexity Analysis}
It is noted that predicting the high dimensional beamforming matrix $\mathbf{V}$ of size $N \times K$ directly using DNN-based solutions \cite{9926157,10146432} will incur high training overhead particularly when the network size becomes large \cite{8935405}. To reduce the training complexity involved, we predict low-dimensional components of the beamforming vectors using the proposed M-ML framework with low training overhead instead of predicting beamforming matrix $\mathbf{V}$. 
We note that the dimension of the predicted variables is reduced from $2(N\times K)$ to $2K$.
% On the other hand, the computational cost incurred for updating the memory is calculated as follows. For the proposed loss sensitivity-based memory updating strategy, the computational cost is incurred mainly by ranking the losses of model during the testing phase in (\ref{eq:012}), thus yielding a computational complexity of $\mathcal{O}(N_n\textrm{log}(N_n))$ with $N_n$ being the number of samples in the memory set $\mathcal{M}^{(i)}_{t}$.
% The $k$-NN based memory updating strategy will incur a computational cost of $\mathcal{O}(N_{m}^{2})$, where $N_{m}$ denotes the number of samples in the query record set $\mathcal{D}_q$. 

On the other hand, the major computational complexity of the M-ML framework is incurred by i) inner model training, ii) outer model testing, iii) memory updating and beamforming reconstruction. In addition, the computational cost incurred for updating the memory is calculated as follows. For the proposed loss sensitivity-based memory updating strategy, the computational cost is incurred mainly by ranking the losses of model during the testing phase in Eq. (\ref{eq:012}), thus yielding a computational complexity of $\mathcal{O}(N_n\textrm{log}(N_n))$ with $N_n$ being the number of samples in the memory set $\mathcal{M}^{(i)}_{t}$. The $k$-NN based memory updating strategy will incur a computational cost of $\mathcal{O}(N_m^2)$. Therefore, the overall complexity involved in the training phase of the proposed M-ML framework is 
$\mathcal{O}( (N_s+N_q) \cdot L \cdot N_{e}^2)$ for each training task, with $L$ and $N_e$ being the total number of layers and number of neurons per layer. Besides, the total complexity involved in the testing phase of the proposed M-ML framework associated with the loss sensitivity-based memory updating strategy is $\mathcal{O}(N_n\textrm{log}(N_n)+ KN)$, whereas the total complexity involved in the testing phase of the proposed M-ML framework associated with $k$-NN based memory updating strategy will incur a computational cost of $\mathcal{O}(N_m^2)$. For the GN-RML-aided beam prediction approach, the complexity of computing $\mathbf{g}\odot \mathbf{n}_{g}$is $\mathcal{O}(N_d)$, with $N_d$ being the number of model parameters. Hence, the total complexity involved in the testing phase of the \textbf{Algorithm \ref{alg:Framwork4}} is $\mathcal{O}(N_d+KN)$.

%------------------> Alg. 3
\alglanguage{pseudocode}
\begin{algorithm}[!t]
\caption{\small Proposed GN-RML Aided Beam Prediction Approach}
\label{alg:Framwork4}
\begin{algorithmic}[1]
\small 
\Require
Initialize the beamforming matrix $\mathbf{V}$, the adaptation learning rate $\boldsymbol{\alpha}$, 
the counter of time slots $t$ = 0.
\Ensure The recovered beamformer $\hat{\bar{\mathbf{V}}}_{}$.
\State \multiline{Decompose the beamforming matrix $\mathbf{V}$ into low-dimensional components $\ddot{\mathbf{w}}$, $\ddot{\mathbf{u}}$, and $\ddot{\mu}$.}
% $\vartriangleright$ \textbf{Low-dimensional component prediction}
\For{{$T$} = 1, 2, $\ldots$}
\State \multiline{Initialize the inner model parameters $\theta_{\ddot{u}}$, $\theta_{\ddot{w}}$, and $\theta_{\ddot{\mu}}$ randomly.}
\For{$t = 1, \ldots, T$} 
\State \multiline{Sample a training task from the set $\mathcal{D}_{s}$ randomly.}
% \State \multiline{Calculate the gradients individually as follows:
% \begin{align}
%     g_{\ddot{u}} \leftarrow \nabla_{\theta_{\ddot{u}}}\sum\nolimits_{i=1}^{N_t}\sum\nolimits_{d=1}^{N_q}Loss(\mathbf{V}_d^{t,i}),
% \end{align}
% \begin{align}
%     g_{\ddot{w}} \leftarrow \nabla_{\theta_{\ddot{w}}}\sum\nolimits_{i=1}^{N_t}\sum\nolimits_{d=1}^{N_q}Loss(\mathbf{V}_d^{t,i}),
% \end{align}
% \begin{align}
%     g_{\ddot{\mu}} \leftarrow \nabla_{\theta_{\ddot{\mu}}}\sum\nolimits_{i=1}^{N_t}\sum\nolimits_{d=1}^{N_q}Loss(\mathbf{V}_d^{t,i}).
% \end{align}
% }
\State \multiline{Calculate the gradients $\hat{\mathbf{g}}$ 
w.r.t. the inner model parameters $\theta_{\ddot{u}}$, $\theta_{\ddot{w}}$, and $\theta_{\ddot{\mu}}$
% $\hat{\mathbf{g}}_{\ddot{u}}$, $\hat{{g}}_{\ddot{w}}$, and $\hat{{g}}_{\ddot{\mu}}$ 
individually using the meta Gaussian DropGrad method via formula (\ref{eq:0027}). 
}
\State \multiline{Update the inner model  parameters $\hat{\theta}_{\ddot{u}}$, $\hat{\theta}_{\ddot{w}}$, and $\hat{\theta}_{\ddot{\mu}}$ individually according to formula (\ref{eq:0025}). 
}
\EndFor
\State \multiline{Update the outer model parameters  $\hat{\boldsymbol{\varphi}}_{\ddot{u}}$, $\hat{\boldsymbol{\varphi}}_{\ddot{w}}$, and $\hat{\boldsymbol{\varphi}}_{\ddot{\mu}}$
by conducting steps 11-12 in \textbf{Algorithm 1}. 
}
\State \multiline{Normalize
the recovered beamformer to satisfy the total power requirement in (\ref{eq:003}) via formula (\ref{equ:04091421}).
% $
% \small
% \label{equ:040914}
% \hat{\bar{\mathbf{V}}}_{t} \leftarrow \frac{\hat{\bar{\mathbf{V}}}_t}{\|\hat{\bar{\mathbf{V}}}_t\|_{2}} \sqrt{P}.
% $
}
\EndFor
\end{algorithmic}
\end{algorithm}
%-------------------------------->

\section{Simulation Results}
\subsection{Simulation Specification}
In this Section, our experiments are conducted in a MISO-BC system adopting the proposed M-ML aided beam prediction method, the proposed GN-RML aided beam prediction method, as well as the following selected counterpart approaches:
\begin{itemize}
    \item \textbf{Traditional meta learning} (i.e., MAML \cite{10096315}): MAML trains an initialized model parameter using iterative gradient updates to achieve good generalization on a new task. MAML updates the network parameters of each task based on the relevant support set $\mathcal{D}_s$ by using (\ref{eq:005}), and updates the global network parameters of all tasks based on the query set $\mathcal{D}_q$ by using (\ref{eq:006}); 
    \item \textbf{Unsupervised learning}: Upon processing a batch of channel realizations, such method updates the randomly initialized model parameters $\varphi_u$ in (\ref{eq:016}), $\varphi_w$ in (\ref{eq:017}), and $\varphi_{\mu}$ in (\ref{eq:018}) in an unsupervised manner for each epoch. Subsequently, the prediction results output from these models undergo the beamformer reconstruction and normalization process as employed in \textbf{Algorithm 1};     
    \item \textbf{Meta learning without pre-training}: It conducts a set of training tasks, each of which involves prediction from a set of channel realizations, without a good meta-initialization;
    \item \textbf{WMMSE} \cite{5756489}: This method iteratively optimizes the 
    low-dimensional components $\mathbf{w}$ in (\ref{eq:0009}), $\mathbf{u}$ in (\ref{eq:0010}), and $\mu$ of the beamforming matrix in an alternative manner by exploiting local channel information, followed by reconstructing and normalizing the beamformer. It converges until a preset error criterion is reached.
\item \textbf{Learning-aided gradient descent} (i.e., LAGD\cite{9805773}): LAGD  optimizes the transmit precoder through implicit GD based iterations, at each of which the optimization strategy is determined
by a neural network, and thus, is dynamic and adaptive.
\end{itemize} 
In addition, we adopt the instantaneous WSR as the performance metric.
The MISO-BC system configuration used in our experiments consists of one BS and three users with the following parameters: $N$ = 3, and $K$ = 3. Besides, $N_s$ = $N_q$ = $N_t$ = 40 are set during the experiments. We adopt the \textit{Adam} optimizer for optimization, and we set the learning rate to be $\alpha$ = 0.01 and $\beta$ = 0.001. For each neural network considered in this paper, the number of neurons in each layer (except for the output layer) is set to be 64.
Besides, for each user the weight $\alpha_k$ is set to be 1. 
Note that, we apply the GN-RML-aided beam prediction method with the dropout rate $P_g$ = 0.5, and $\mathrm{Var}(\mathbf{n}_g)$ = 1 in the following experiments.
Besides, we use PyTorch 2.1.1 with an NVIDIA RTX 3090 GPU to conduct the following experiments.

\begin{figure}[t!]
    \begin{center}
    \includegraphics[width=0.47\textwidth]{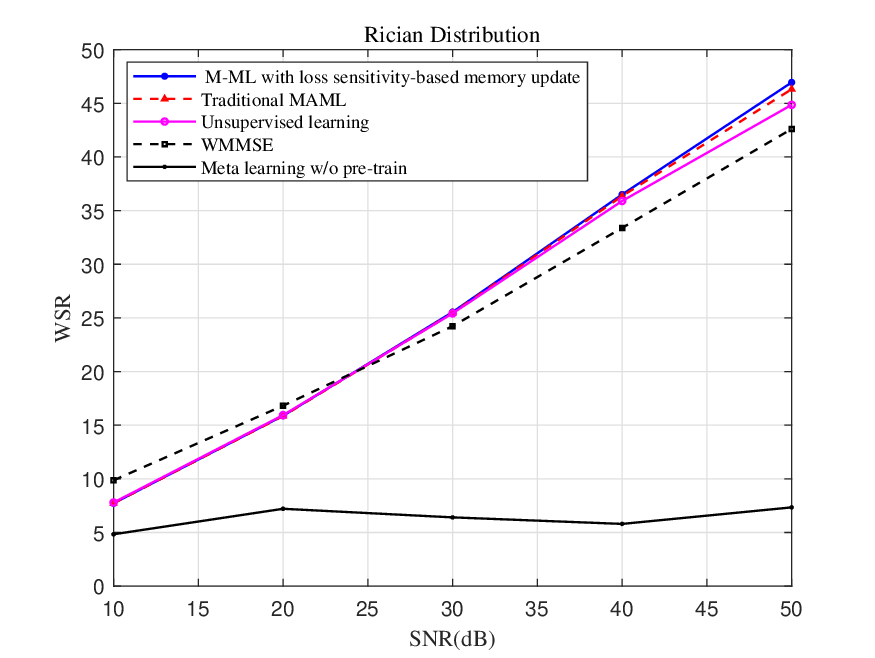}
    \end{center}
 \caption{WSR performance comparison of distinct beam prediction approaches in Rician fading environment when $N$ = 3, $K$ = 3, and $N_t$ = $N_s$ = $N_q$ = 40.}    \label{fig:5}
\end{figure}

It is noted that both i) Rayleigh, ii) Rician and iii) Nakagami-$m$ fading channel environments are considered and used for training. During the training of the experiments, the proportion of each of the two different types of channels is kept equal to 50$\%$ and the dataset with the size of 500 is trained using each learning method. 
Notice that,
in our experiments we separately verify the adaptability of the proposed framework in two different scenarios:
\begin{itemize}
    \item \textit{Scene 1}: In the first scene, the environment where both training and testing channels follow the same distribution. In the following simulations we adopt the Rician channel environment for model training and testing;
\item \textit{Scene 2}:
In the second scene, the environment of training channel is misaligned with the environment of the testing channel. In this scene, we adopt Rician channel environment for model training, whereas using the Nakagami-$m$ environment for model testing.
\end{itemize}
% In addition, we set the bandwidth to be 0.05 GHz, the transmitting antennas to be 32, the number of subcarries to be 32, and the number of
% paths to be 25. 
% Except for the special mentioned cases, we use the first 10000 CSI data in \textit{dense area} (\textit{sparse area}) as the dataset, with 5000 for training and 5000 for testing. Moreover, the data are normalized to the range of [-1,1]. 

\begin{figure}[t!]
    \begin{center}
    \includegraphics[width=0.47\textwidth]{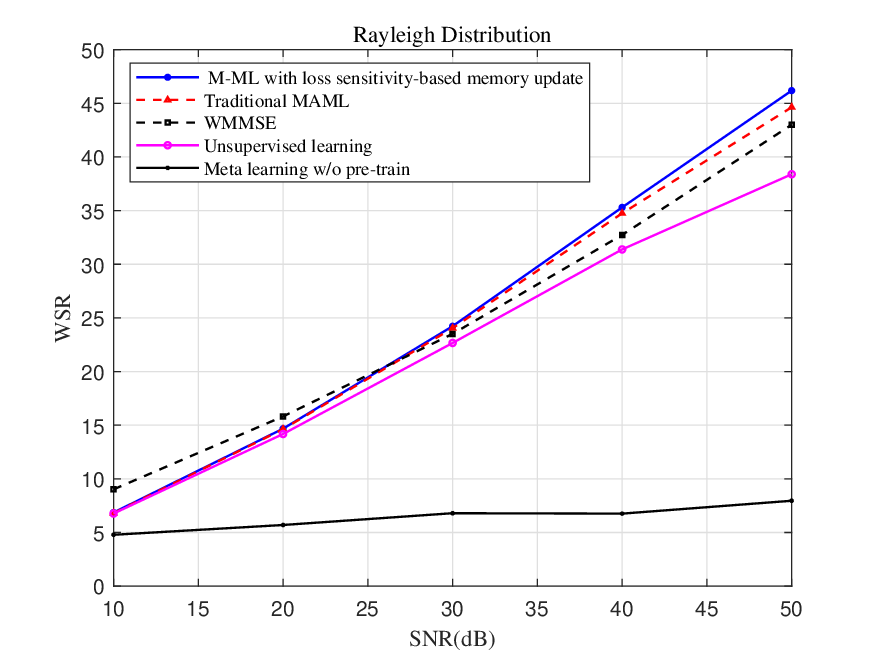}
    \end{center}
 \caption{WSR performance comparison of distinct beam prediction approaches in Rayleigh fading environment when $N$ = 3, $K$ = 3, and $N_t$ = $N_s$ = $N_q$ = 40.}    \label{fig:6}
\end{figure}

\subsection{WSR Performance Comparisons}
\subsubsection{M-ML Aided Beam Prediction Performance}
%
% For the memory-based meta-learning beam prediction in the model, during the experiments, we performed comparison tests by four methods: unsupervised learning, traditional meta-learning, WMMSE and no pre-training and only adding memristors.

% 
In Figs. \ref{fig:5} and \ref{fig:6}, we first test the model in the same distribution as the training environment so as to show the performance of the model in the test environment. It is easy to see from both Figs. \ref{fig:5} and \ref{fig:6} that, the traditional MAML approach outperforms the WMMSE and unsupervised learning approaches at high signal-to-noise ratios (SNRs). The relevant reasons lie in that: i) as at the high SNR regions, the nonconvexity in the WSR maximization problem is amplified. M-ML has the capability of searching the global optimal point along the complex geometry, while the other algorithms might search the local optimal points in the geometry surface of the WSR optimization problem. ii) However, at low SNR regions, WMMSE might easily search the global optimal point due to the shrank solution space. Besides, it resolves the non-convex WSR maximization problem with higher convergence.
In addition, our M-ML introduces a memorization mechanism based on the traditional MAML, which is designed to enhance the model's ability of rapid adaptation to i) the environment where both training and testing channels follow the same distribution and ii) the new environments. 
Although the performance improvement is not significant in the test environment with the same distribution, this indicates potential room for improvement in the model's adaptability when encountering different environments.

\begin{figure}[t!]
    \begin{center}
    \includegraphics[width=0.47\textwidth]{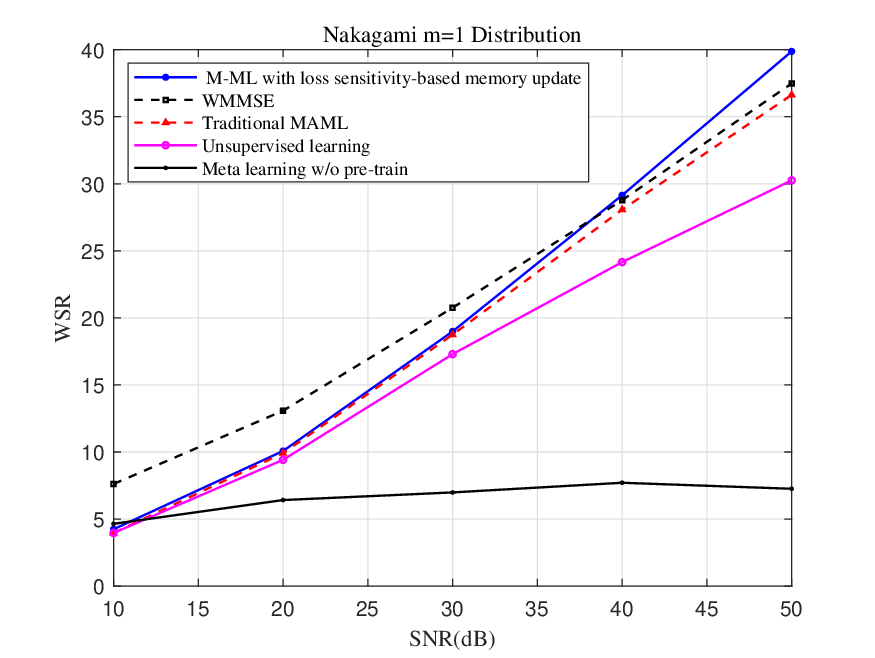}
    \end{center}
 \caption{WSR performance comparison of distinct beam prediction approaches in Nakagami-$m$ fading environment when $N$ = 3, $K$ = 3, $N_t$ = $N_s$ = $N_q$ = 40, and $m$ = 1.}    \label{fig:7}
\end{figure}

% \begin{figure}[t!]
% % \setlength{\abovecaptionskip}{-0.2cm}
% % \setlength{\belowcaptionskip}{-0.5cm}
%     \begin{center}
%     \includegraphics[width=9cm]{fig/rician1.eps}
%     \end{center}
%  \caption{Performance of memory-based meta-learning beam prediction in an rician environment}    \label{fig:5}
% \end{figure}

% \begin{figure}[t!]
% % \setlength{\abovecaptionskip}{-0.1cm}
% % \setlength{\belowcaptionskip}{-0.5cm}
%     \begin{center}
%     \includegraphics[width=0.47\textwidth]{fig/nami10d2.eps}
%     \end{center}
%  \caption{WSR performance comparison of distinct beam prediction approaches in Nakagami-$m$ fading environment when $N$ = 3, $K$ = 3, $N_t$ = $N_s$ = $N_q$ = 40, and $m$ = 10.}    \label{fig:8}
% \end{figure}
\begin{figure}[t!]
    \begin{center}
    \includegraphics[width=0.47\textwidth]{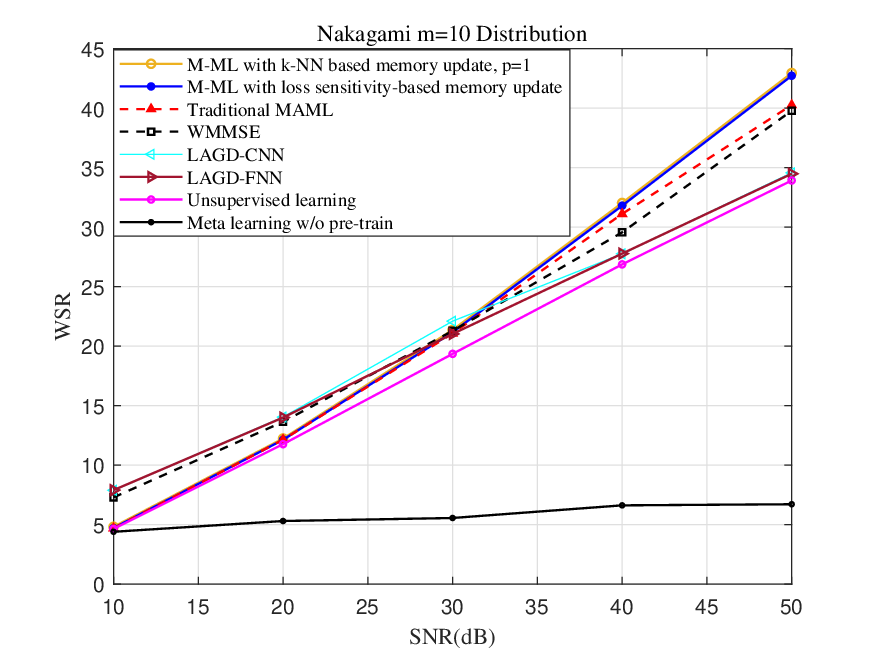}
    \end{center}
 \caption{WSR performance comparison of distinct beam prediction approaches in Nakagami-$m$ fading environment when $N$ = 3, $K$ = 3, $N_t$ = $N_s$ = $N_q$ = 40, and $m$ = 10.}    \label{fig:8}
\end{figure}

\begin{figure}[t!]
    \begin{center}
    \includegraphics[width=0.47\textwidth]{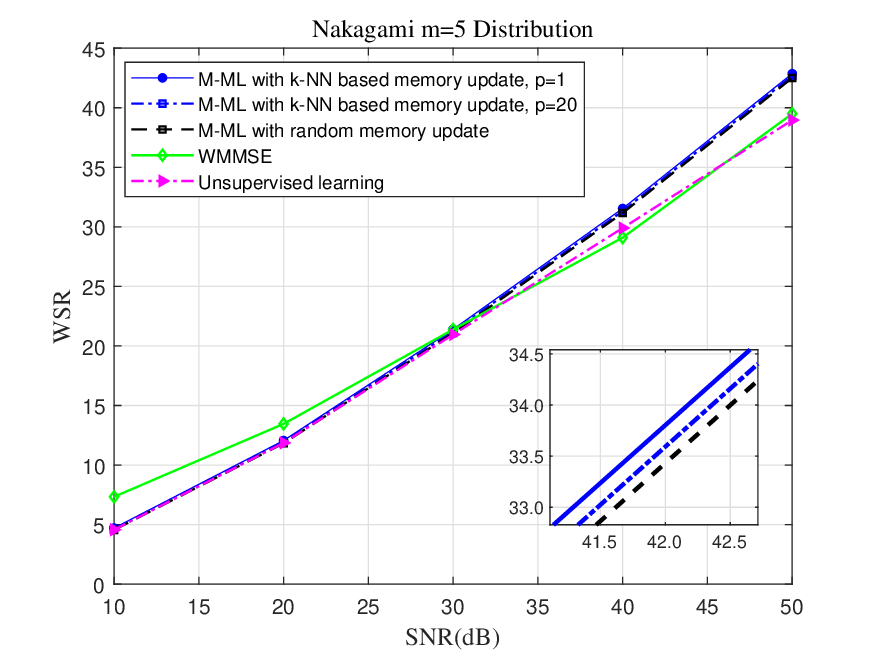}
    \end{center}
 \caption{WSR performance comparison of distinct beam prediction approaches in Nakagami-$m$ fading environment when $N$ = 3, $K$ = 3, and $N_t$ = $N_s$ = $N_q$ = 40, $m$ = 5, and $p=\{1, 20\}$.}    \label{fig:11}
\end{figure}

% \begin{figure}[t!]
% % \setlength{\abovecaptionskip}{-0.1cm}
% % \setlength{\belowcaptionskip}{-0.5cm}
%     \begin{center}
%     \includegraphics[width=0.47\textwidth]{fig/WSR_Convergence_20dB_01.eps}
%     \end{center}
%  \caption{WSR performance comparison of distinct beam prediction approaches in different scenes when $N$ = 3, $K$ = 3, $N_t$ = $N_s$ = $N_q$ = 40, and SNR = 20 dB.}    \label{fig:011}
% \end{figure}
\begin{figure}[t!]
    \begin{center}
    \includegraphics[width=0.47\textwidth]{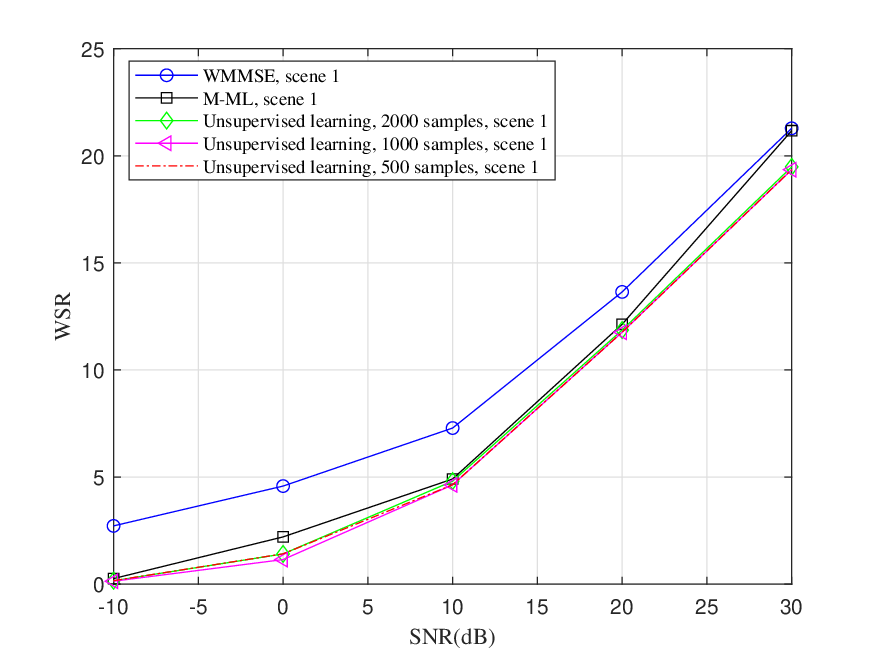}
    \end{center}
 \caption{WSR performance comparison of distinct beam prediction approaches in scene 1  when $N$ = 16 and $K$ = 16.}    \label{fig:011}
\end{figure}

% \begin{figure}[t!]
% % \setlength{\abovecaptionskip}{-0.1cm}
% % \setlength{\belowcaptionskip}{-0.5cm}
%     \begin{center}
%     \includegraphics[width=0.47\textwidth]{fig/WSR_Convergence_40dB_01.eps}
%     \end{center}
%  \caption{WSR performance comparison of distinct beam prediction approaches in distinct scenes when $N$ = 3, $K$ = 3, $N_t$ = $N_s$ = $N_q$ = 40, and SNR = 40 dB.}    \label{fig:012}
% \end{figure}
\begin{figure}[t!]
    \begin{center}
    \includegraphics[width=0.47\textwidth]{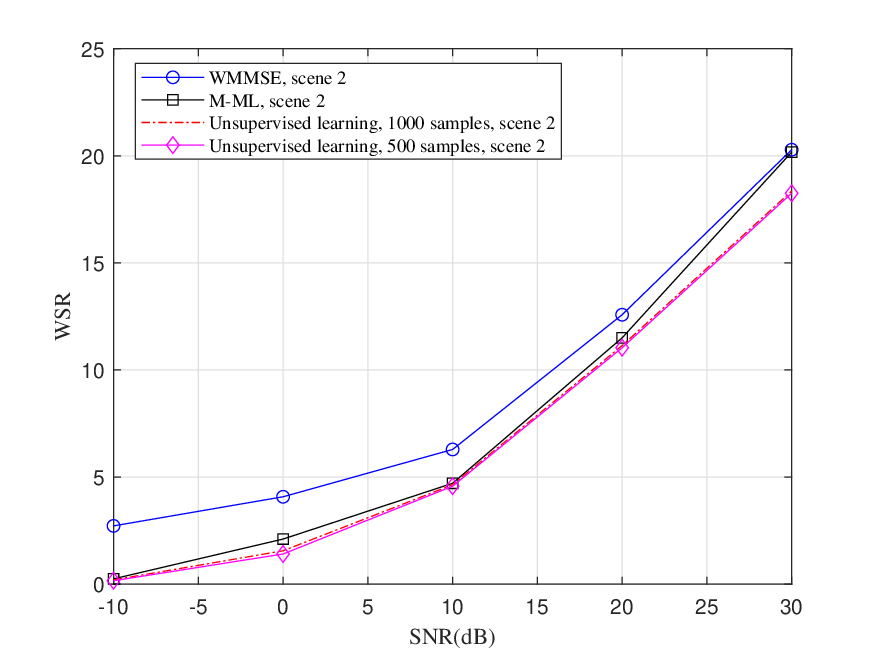}
    \end{center}
 \caption{WSR performance comparison of distinct beam prediction approaches in scene 2 when $N$ = 16 and $K$ = 16.}    \label{fig:012}
\end{figure}

Subsequently, in Figs. \ref{fig:7} and \ref{fig:8}, we extend the test range by choosing a test environment with a different distribution from the training environment, which is Nakagami-$m$ fading channel distribution with $m$=1 and $m$=10, to further evaluate the model's generalization ability. Simulation results show that all of the above mentioned methods suffer from WSR performance degradation when the models encounter mismatched training and testing channel distributions. In particular, the unsupervised learning method performs the worst performance in this case due to their lack of exploiting prior knowledge of the environment. However, we observe that the proposed M-ML scheme significantly improves the model's adaptability and WSR performance in new environments. This is due to the loss sensitivity and $k$-NN inspired memory set selection mechanisms, which are helpful to store key data from memory and the latest data collected from the new environments. In addition, the proposed $k$-NN-based memory set selection mechanism with $p$=1 achieves a similar WSR performance when comparing with the loss sensitivity-based memory set selection scheme under channel distribution shifts. However, the latter scheme incurs a smaller complexity cost than the former one. 
The benefit of the improved adaptability of M-ML is not only reflected in the fast response to new environments, but also means that the model is able to maintain stable and reliable performance in a wider range of application scenarios. In addition,
at high SNR region, M-ML has the capability of searching the global optimal point in the geometry surface of the WSR optimization problem. The other methods might search the local optimal points along the complex geometry. 
% At low SNR region, WMMSE might easily search the global optimal point and solve the WSR maximization problem with higher convergence. 
Based on the above results, we deduce that M-ML can provide strong support for the robustness and flexibility of ML in real-world applications.
We observe from Fig. \ref{fig:5} that the performance improvement against the traditional MAML in Scene 1 is not significant. However, we can see from Fig. \ref{fig:7} that, the performance gaps between the proposed M-ML-based method and the traditional MAML-based scheme in Scene 2 is enlarged. This is because at high SNR regions, the nonconvexity in the WSR maximization problem is amplified. One of the main drawbacks of the MAML algorithm is that it is complicated in the training and testing phases. Furthermore, when encountering divergent training and testing channel distributions, MAML’s adaptation process will become to be slow, thus severely degrading the accuracy performance.

Fig. \ref{fig:11} portrays the WSR performance of distinct beam prediction approaches in Nakagami-$m$ fading environment with $m$ = 5. Fig. \ref{fig:11} reveals that the proposed M-ML with $k$-NN based memory updating strategy outperforms the state-of-the-art WMMSE approach in high-SNR regimes. In addition, the proposed approach can achieve superior WSR performance than the unsupervised learning. This is due to the fact that the proposed $k$-NN based memory updating method can identify the old samples having the largest dissimilarity to new channel realization samples, thus forming memory set $\hat{\bar{\mathcal{M}}}_{t}$ that assists model to learn more robust and generalizable features. Besides, Fig. \ref{fig:11} shows that enlarging the value of parameter $m$ results in degraded WSR performance. However, it still surpasses the unsupervised learning and the WMMSE methods in high-SNR regimes.

In Figs. \ref{fig:011} and \ref{fig:012}, we present the obtained WSR results of the M-ML, full iterative WMMSE, and unsupervised learning algorithms averaged over 1000 channel realizations in two different scenes, using larger settings of $N$ and $K$. Fig. \ref{fig:011}  illustrates that the full iterative WMMSE algorithm achieves higher WSR performance than the proposed M-ML approach at low-SNR regimes. However, the performance gap between M-ML and WMMSE decreases as increasing SNR. Besides, the performance gap between M-ML and unsupervised learning increases at high-SNR regimes. Fig. \ref{fig:012} shows that the performance benefit of the proposed M-ML approach still holds even under channel distribution shifts due to benefits brought by the memory update mechanism. Besides, when comparing Fig. \ref{fig:011} with Fig. \ref{fig:012}, we observe that: i) all these schemes suffer from performance degradation in Scene 2. However, the performance gap between M-ML and WMMSE becomes smaller at low-SNR regimes, due to the high model adaptability performance in new environments. ii) M-ML exhibits strong scalability and robustness in realistic mmWave massive MIMO, even for large settings of $N$ and $K$.

% In Figs. \ref{fig:011} and \ref{fig:012}, we present the obtained WSR results of the M-ML and WMMSE algorithms averaged over 1000 channel realizations in two different scenes. Fig. \ref{fig:011} shows that M-ML enables higher WSR performance than WMMSE in Scene 1. However, WMMSE surpasses the proposed M-ML approach when $\text{SNR}$ = 20 dB. Besides, the performance gap between the above two methods decreases as increasing the round of iterations. Fig. \ref{fig:012} illustrates that M-ML outperforms WMMSE when $\text{SNR}$  = 40 dB even in scene 2. Besides, when comparing Fig. \ref{fig:011} with Fig. \ref{fig:012}, we observe that: i) our M-ML framework is feasibility in Scene 1 in low-SNR regimes when it converges; ii) as for the Scene 2, the WSR performance gap of the above-mentioned methods decreases when increasing the value of SNRs. This implies that our framework is valid to the Scene 2 in high-SNR regimes.

\subsubsection{GN-RML Aided Beam Prediction Performance}
For the experimental process of ML regularization, we focus on the impact of adding a regularization method to the traditional ML in unseen environments on the results of WSR for model training.
We evaluate the WSR performance using the following three approaches: i) traditional ML (i.e., MAML \cite{Finn2017ModelAgnosticMF}), ii) regularization without noise (regularization) and iii) regularized ML with Gaussian noise (Gaussian regularization).

\begin{figure}[t!]
    \begin{center}
    \includegraphics[width=0.47\textwidth]{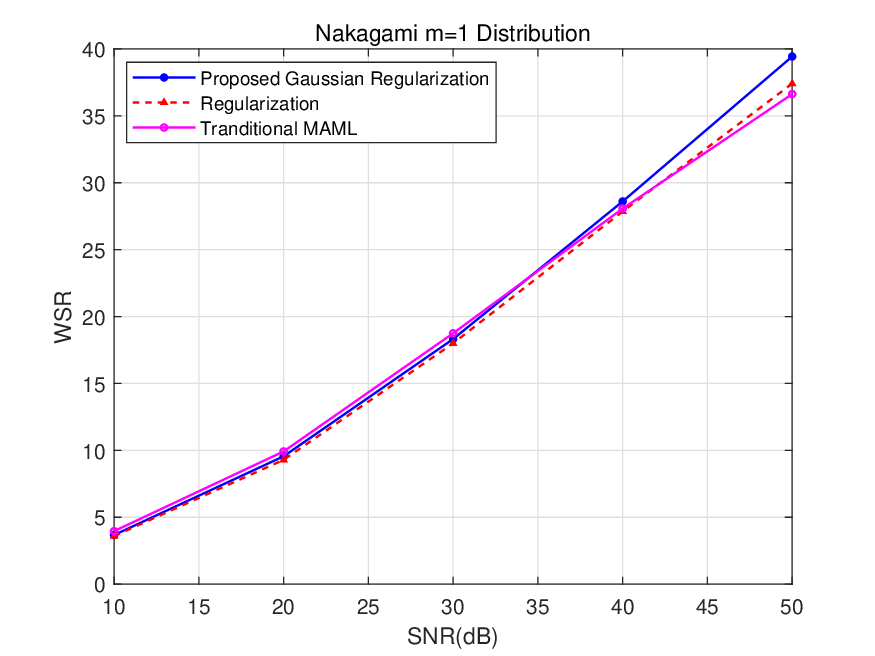}
    \end{center}
 \caption{WSR performance comparison of the proposed Gaussian regularization and the regularization without noise, as well as the traditional MAML methods in Nakagami-$m$ fading environment when $N$ = 3, $K$ = 3, $N_t$ = $N_s$ = $N_q$ = 40, $m$ = 1.}    \label{fig:3}
\end{figure}

\begin{figure}[t!]
    \begin{center}
    \includegraphics[width=0.47\textwidth]{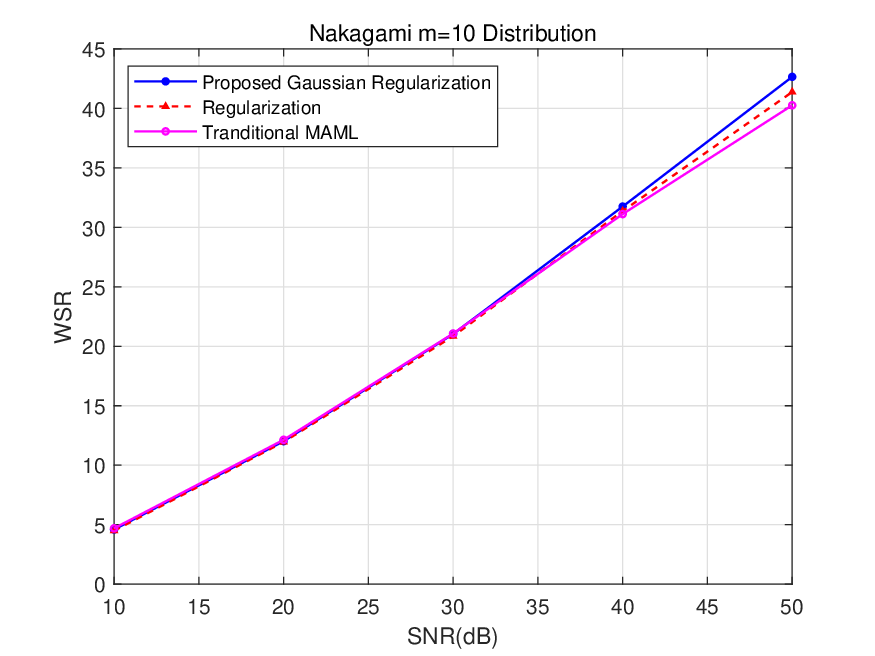}
    \end{center}
 \caption{WSR performance comparison of the proposed Gaussian regularization and the regularization without noise, as well as the traditional MAML methods in Nakagami-$m$ fading environment when $N$ = 3, $K$ = 3, $N_t$ = $N_s$ = $N_q$ = 40, and $m$ = 10.}    \label{fig:4}
\end{figure}

As can be seen from Figs. \ref{fig:3} and \ref{fig:4}, the proposed GN-RML aided beam prediction approach can achieve superior WSR performance than the traditional MAML method. The related reason behind is that weight regularization is a strategy that enhances the generalization ability of a model by limiting the complexity of its parameters, and its main advantages include reducing meta-overfitting and improving the stability and generalization ability of the model. The proposed approach provides limited performance gain in relative to the traditional MAML scheme at low-SNR regimes, as it might search the local optimal points in the geometry surface of the WSR optimization problem.
However, weighted regularization ML still suffers from the problem that the regularization parameters need to be carefully adjusted to achieve the best results and may oversimplify the model in some cases. Therefore, we use regularized ML with the introduction of Gaussian noise, which, as can be seen in the figure, provides the best performance. The main reason for this is its ability to effectively model the uncertainty in the training data, which motivates the model to learn a more robust feature representation and reduce the sensitivity to noise and outliers in the training data. In addition, Gaussian noise can change the decision boundary of the model in a smooth way, which helps to avoid the model from falling into a local optimum, thus improving its generalization performance on unknown data. Meanwhile, due to the mathematically excellent properties of the Gaussian distribution, Gaussian noise regularization is easy to implement in practical applications, and the intensity of regularization can be controlled by adjusting the variance of the noise, providing a flexible regularization strategy for the model.

\section{Concluding Remarks}
In this paper, we proposed an innovative M-ML framework for predicting mmWave beam, aiming to significantly improve the model's adaptability in new environments. Our method extends the traditional MAML framework by introducing a core component, i.e., the memristor, which enhances the model's ability of dynamically learning and memorizing the features of new environments.
Subsequently, we presented an innovative 
GN-RML aided beam prediction strategy. This strategy successfully simulates the uncertainty in the training data by introducing Gaussian noise in the training phase, thus driving the model to learn a more robust feature representation. This strategy not only enhances the model's resistance to input perturbations, but also improves its stability and accuracy in complex environments.
Simulation results demonstrated that, our approach achieves significant performance gains in trained scenarios. By dynamically storing the data collected in the new environment into memory and fully utilizing them during subsequent training and testing, our model can swiftly adapt to changes in the new environment and achieve more accurate predictions. In future work, we will further study the sensitivity of the memory set size $M$ to channel non-stationarity and identify the preferred memory set size $M$ for different task sizes across distinct scenes.

\bibliographystyle{IEEEtran}%
\bibliography{ref1}

\end{document}